\documentstyle[twocolumn,prb,aps,epsf]{revtex}

\preprint{NUC-MINN-01/...-T}

\newcommand{\be}{\begin{equation}}
\newcommand{\ee}{\end{equation}}

\newcommand{\ba}{\begin{eqnarray}}
\newcommand{\ea}{\end{eqnarray}}
\newcommand{\A}{\overcirc{A}}

\begin{document}

\draft

\title{Magnetic field driven metal-insulator phase transition in 
planar systems}

\author{E.V. Gorbar$^{*}$}
\address{Universidade Federal de Juiz de Fora, 
Juiz de Fora 36036-330, Brazil}

\author{V. P. Gusynin}
\address{Bogolyubov Institute for Theoretical Physics,
        03143, Kiev, Ukraine\\
    and Department of Physics, Nagoya University,
        Nagoya 464-8602, Japan}

\author{V.A.~Miransky$^{*}$}
\address{Department of Applied Mathematics, University of Western
Ontario, London, Ontario N6A 5B7, Canada}

\author{I. A. Shovkovy$^{*}$}
\address{School of Physics and Astronomy, University of Minnesota,
        Minneapolis, MN 55455, USA}

\date{March 12, 2002}
\maketitle

\begin{abstract} 
A theory of the magnetic field driven (semi-)metal-insulator phase
transition is developed for planar systems with a low density of carriers
and a linear (i.e., relativistic like) dispersion relation for low energy
quasiparticles. The general structure of the phase diagram of the theory
with respect to the coupling constant, the chemical potential and
temperature is derived in two cases, with and without an external magnetic
field. The conductivity and resistivity as functions of temperature and
magnetic field are studied in detail. An exact relation for the value of
the ``offset" magnetic field $B_c$, determining the threshold for the
realization of the phase transition at zero temperature, is established.
The theory is applied to the description of a recently observed phase
transition induced by a magnetic field in highly oriented pyrolytic
graphite.  
\end{abstract} 


\section{Introduction}
\label{Introduction}

Although during recent years there has been important progress in
understanding non-Fermi liquid dynamics in dimensions $D>1$, an 
understanding of them is still very far from being complete. It is 
rather clear that non-Fermi liquid behavior yields examples of 
sophisticated nonperturbative dynamics which should be described by 
advanced methods of quantum field theory.

It was recognized rather long ago that relativistic field models can
serve as effective theories for the description of long wavelength
excitations in condensed matter systems (for a review, see
Ref.~\onlinecite{Fra}). In particular, they can be applied to a wide class
of (quasi-)planar systems. In this case, the corresponding relativistic
theories are $(2+1)$-dimensional, i.e., they are formulated in 
$(2+1)$-dimensional Minkowski space with two space like and one time like
coordinates. It is important that amongst these condensed matter systems
are such as high-$T_c$ superconductors and carbon-based materials (for a
list of papers using relativistic field approach to these systems see
Refs.~\onlinecite{highTc,kim,highTc2,thermo,Semenoff,carbon,Khvesh,Khvesh2}).

In this paper, we will develop a consistent approach to studying these
systems by making use of so called reduced $(3+1)$-dimensional gauge
theories.\cite{GGM,AFKM} These theories will share the following common
feature. Their gauge fields (e.g., the electromagnetic field) responsible
for interparticle interaction would be able to propagate in a 
three-dimensional bulk, while fermion fields (e.g., describing electron-
and hole-type quasiparticles) would be localized on two-dimensional planes. 
A typical example of a condensed matter system of this type is
graphite. It has been known for a long time that fermionic quasiparticles
in graphite are nearly two dimensional.\cite{Wallace} In addition, graphite is
a semimetal whose low-energy quasiparticles have nearly linear dispersion
law (just like massless relativistic 
particles).\cite{Wallace,McClure1,Semenoff} The Coulomb interaction 
between quasiparticles is provided by gauge fields which, unlike the
quasiparticles themselves, are three dimensional in nature.

Recently, the dynamics of reduced QED was studied in
Refs.~\onlinecite{GGM,AFKM}. In those papers, purely relativistic theories
were considered: in particular the velocities of both massless fermions
and photons were equal to the speed of light $c$. In realistic condensed
matter systems, the Fermi velocity of gapless fermions $v_F$ is of course
much less than $c$. This in turn implies that the static Coulomb forces
provide the dominant interactions of fermions. This feature makes quite a
difference in the analysis.

In this paper we will describe such ``realistic" reduced gauge theories
with and without an external magnetic field perpendicular to the basal
plane. We are particularly interested in the possibility of a spontaneous
generation of a gap in the one-quasiparticle spectrum. This might be
viewed as a (semi)metal-insulator phase transition. The influence of the
magnetic field, as would become clear in a moment, is very powerful in
driving (or ``catalyzing") this type of transitions.

The phenomenon of the magnetic catalysis of dynamical symmetry breaking
was established as a universal phenomenon in a wide class of $(2+1)$- and
$(3+1)$-dimensional relativistic models in Refs.~\onlinecite{1,2} (for
earlier consideration of dynamical symmetry breaking in a magnetic field
see Refs.~\onlinecite{Kawati,Klim}).

The general result states that a constant magnetic field leads to the
generation of a fermion dynamical mass (a gap in a one-particle energy
spectrum) even at the weakest attractive interaction between fermions. The
essence of this effect is the dimensional reduction $D\to D-2$ in the
dynamics of fermion pairing in a magnetic field. At weak coupling, this
dynamics is dominated by the lowest Landau level (LLL) which is
essentially $(D-2)$-dimensional.\cite{1,2} The applications of this effect
have been considered both in condensed matter physics\cite{thermo,Khvesh}
and cosmology (for reviews see Ref.~\onlinecite{reviews}).

The main motivation of the present study was the experimental data
reported in Refs.~\onlinecite{Exp1,SSCom115,Exp2} and their interpretation
(based on the phenomenon of the magnetic catalysis) suggested in
Ref.~\onlinecite{Khvesh}. It was observed in those experiments that
samples of highly oriented pyrolytic graphite in an external magnetic
field show a qualitative change of their resistivity as a function of
temperature, that was interpreted as a metal-insulator phase transition.
The effect is clearly seen only for a magnetic field perpendicular to the
basal plane, suggesting that the orbital motion of quasiparticles is
responsible for the change of the conductivity dependence.

The suggestion of Ref.~\onlinecite{Khvesh} was that this phenomenon can be
a manifestation of the magnetic catalysis, when a dynamical gap, connected
with a quasiparticle-hole pairing, is generated in a magnetic field. In
this paper, we will develop a detailed theory of the magnetic-field-driven
metal-insulator phase transition in planar systems, based on reduced QED.  
The general structure of the phase diagram of such systems will be
described in two cases, with and without an external magnetic field. The
behavior of the electric conductivity (resistivity) in these systems will
be described in detail. This will allow us to conclude that, in the
presence of a magnetic field, the generation of a dynamical gap in planar
systems can indeed manifest itself as a metal-insulator phase transition
in the behavior of the resistivity $\rho(T,B)$ as a function of the
magnetic field and temperature.

It will be also shown that there exist clearly distinguishable signatures
of different types of the phase transition. While the resistivity
$\rho(T)$ is a smooth function at the critical point $T = T_c$ in the case
of a non-mean-field second-order  
phase transition, there are a discontinuity
and a kink in $\rho(T)$ at $T = T_c$ in the cases of the first-order and
mean-field phase transitions, respectively. The conclusion of the present
analysis concerning the possibility of the realization of the scenario of
the magnetic catalysis in highly oriented pyrolytic graphite is quite
positive.

One of the central results of this paper is an explanation of the
existence of an ``offset" field $B_c$ observed in the 
experiments.\cite{Exp1,SSCom115,Exp2} 
As we will discuss in detail in Sec.~\ref{HOPG},
the value $B_c$ determines the threshold for the generation of a dynamical
gap at {\it zero} temperature: it happens only if $B > B_c$. It is
remarkable that, as will be shown in Sec.~\ref{sec:mag-field}, the
existence of $B_c$ is a robust consequence of the mechanism of the
magnetic catalysis. Moreover, the {\it exact} relation for $B_c$ will be
pointed out. It is: 
\be 
|eB_{c}| = \frac{2\pi c n}{N_{f}}, 
\ee 
where $N_f$
is the number of fermion species (``flavors") and $n$ is a charge density
of carriers ($N_f = 2$ in graphite). While the existence of this exact
relation is noticeable in itself, its experimental verification would be a
critical check of the validity of the magnetic catalysis scenario in
highly oriented pyrolytic graphite.

The paper is organized as follows. In Sec.~\ref{sec:model} general
features of the model (reduced QED) are described. In Sec.~\ref{gapequat}
we analyze the gap equation and establish the phase diagram in reduced QED
without magnetic field. In Sec.~\ref{sec:mag-field} the gap equation in
reduced QED with an external magnetic field is studied. The resistivity
and conductivity in this system are studied in detail in
Sec.~\ref{condres}. Sec.~\ref{HOPG} is devoted to the interpretation of
the experimental data in highly oriented pyrolytic graphite. In
Sec.~\ref{conclusion}, we summarize the results of this work. There are
also three Appendices. The symmetry of $(2+1)$-dimensional fermions is
considered in Appendix~\ref{Appfer}. A derivation of the polarization
function and the gap equation in reduced QED is done in
Appendix~\ref{AppA}. In Appendix~\ref{AppB}, the effective potential for
reduced QED with a nonzero chemical potential is derived.

\section{Model}
\label{sec:model}

In this section, we describe the general features of the model. As
mentioned in Sec.~\ref{Introduction}, 
the main assumption of the reduced dynamics of
the planar systems is that the fermionic quasiparticles are confined to a
plane, while the gauge fields are free to propagate in the 
three-dimensional bulk.

\begin{figure}
\epsfxsize=8.0cm
\epsffile[200 60 485 220]{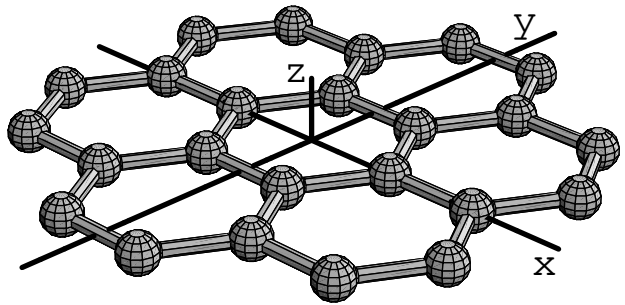}
\caption{The schematic lattice structure of a single layer of 
graphite.}
\label{fig:1}
\end{figure}

A similar setting was recently studied in a class of relativistic models
in Refs.~\onlinecite{GGM,AFKM}. Here, however, we consider a strongly
nonrelativistic model (with the Fermi velocity $v_{F}$ being much less
than the speed of light) which could be applied to realistic planar
condensed matter systems such as highly oriented pyrolytic graphite; 
see Fig.~\ref{fig:1}.
The spatial coordinates on the plane (e.g., a single layer of graphite)
are denoted by $\vec{r}=(x,y)$. The orthogonal direction is labeled by the
$z$ coordinate. Thus, the most general bulk spatial vector is given by
$\vec{R}=(x,y,z)$.

The lagrangian density of the electromagnetic field (in the bulk) 
is given by
\ba
{\cal L}_{em} &=& \frac{1}{8\pi} \left(\varepsilon_{0} \vec{E}^{2}
-\frac{1}{\mu_{0}} \vec{B}^{2} \right) 
-A_{0} \rho +\frac{1}{c} \vec{A}\cdot \vec{j},
\label{L-1}
\ea
where $\varepsilon_{0}$ is the dielectric constant, 
$\mu_{0}$ is the magnetic permeability,  
$A_{0}$ and $\vec{A}$ are the scalar and vector potentials. The
electric and magnetic fields are
\ba
\vec{E} &=& - \vec{\nabla} A_{0} -\frac{1}{c} \partial_{t} \vec{A}, \\
\vec{B} &=& \vec{\nabla} \times \vec{A}.
\ea
The interacting terms, with the quasiparticle charge density $\rho$ and
current $\vec{j}$, were also included in the lagrangian density in
Eq.~(\ref{L-1}). Now, the lagrangian density of quasiparticles themselves
(defined only on the plane) reads 
\be
{\cal L}_{0} = v_{F}
\bar{\Psi}(t,\vec{r})
\left(\frac{i\gamma^{0}(\partial_{t}+i\mu)}{v_{F}} 
-i\gamma^{1} \partial_{x}
-i\gamma^{2} \partial_{y}\right)\Psi(t,\vec{r}),
\label{L-free}
\ee
where $\Psi(t,\vec{r})$ is a four-component spinor, $\bar{\Psi}=
\Psi^{\dagger}\gamma^{0}$, and the $4\times 4$ Dirac $\gamma$-matrices
furnish a reducible representation of the Clifford (Dirac) algebra in
$2+1$ dimensions (see Appendix~\ref{Appfer}).\cite{JT,P} In order to
describe the situation with a finite ``residual"  density of carriers,
here the chemical potential $\mu$, connected with the electric charge,
was introduced.

We will consider the case when the fermion fields carry an additional,
``flavor", index $i=1,2,\dots,N_f$ (in the example of graphite, $N_{f}=2$,
see Refs.~\onlinecite{carbon,Khvesh}). Then, the symmetry of the
lagrangian (\ref{L-free}) is $U(2N_{f})$ (see Appendix~\ref{Appfer}).

In the case of minimal coupling of the electromagnetic field, the
quasiparticle charge density and current take the following explicit 
forms:
\ba
\rho(t,\vec{R}) &=& e \bar{\Psi}(t,\vec{r})\gamma^{0}\Psi(t,\vec{r}) 
\delta (z) , \\
j_{x}(t,\vec{R}) &=& ev_{F} \bar{\Psi}(t,\vec{r})\gamma^{1}\Psi(t,\vec{r}) 
\delta (z) , \\
j_{y}(t,\vec{R}) &=& ev_{F} \bar{\Psi}(t,\vec{r})\gamma^{2}\Psi(t,\vec{r}) 
\delta (z) , \\
j_{z}(t,\vec{R}) &=& 0.
\ea
Proceeding as in Ref.~\onlinecite{GGM}, the initial action can be reduced
to the plane. Then, neglecting relativistic corrections of order
$(v_{F}/c)^{2}$, we are left with the following action of
interacting quasiparticles:
\ba
S_{qp} &\simeq& \int dt d^{2} \vec{r} {\cal L}_{0}(t,\vec{r})
-\frac{1}{2} 
\int dt \int dt^{\prime} 
\int d^{2} \vec{r} \int d^{2} \vec{r}^{\prime} 
\nonumber \\
&\times& \bar{\Psi}(t,\vec{r})\gamma^{0}\Psi(t,\vec{r}) 
U_{0}(t-t^{\prime},|\vec{r}-\vec{r}^{\prime}|)
\nonumber \\
&\times& \bar{\Psi}(t^{\prime},\vec{r}^{\prime})\gamma^{0}
\Psi(t^{\prime},\vec{r}^{\prime}).
\label{action}
\ea
The bare potential $U_{0}(t,|\vec{r}|)$ takes the following 
simple form:
\be
U_{0}(t,|\vec{r}|) = \frac{e^2\delta(t)}{\varepsilon_{0}} 
\int \frac{d^{2}\vec{k}}{(2\pi)^{2}}  
\exp(i\vec{k}\cdot\vec{r})\frac{2\pi}{|\vec{k}|}
=\frac{e^{2}\delta(t)}{\varepsilon_{0}|\vec{r}|}.
\ee
Note, however, that in many cases of interest (e.g., in the case of a
finite temperature and/or a finite density and/or a nonzero magnetic
field), the polarization effects may considerably modify this bare Coulomb
potential. Thus, the interaction should rather be given by
\be
U(t,|\vec{r}|) = \frac{e^2}{\varepsilon_{0}}
\int\frac{d\omega}{2\pi} \int \frac{d^{2}\vec{k}}{2\pi}
\frac{\exp(-i\omega t+i\vec{k}\cdot\vec{r})}
{|\vec{k}|+\Pi(\omega,|\vec{k}|)},
\label{retard}
\ee
where the polarization function $\Pi(\omega,|\vec{k}|)$ is 
proportional (with a factor of $2\pi/\varepsilon_{0}$) to
the time component of the photon polarization tensor.

Adding a mass (gap) term $\Delta_{0}\bar\psi\psi$ into the action
(\ref{action}) would reduce the $U(2N_{f})$ symmetry down to the
$U(N_f)\times U(N_f)$ (see Appendix~\ref{Appfer}). Therefore the dynamical
generation of a fermion gap (connected with a quasiparticle-hole pairing)
will lead to the spontaneous breakdown of the $U(2N_f)$ down to the
$U(N_f)\times U(N_f)$.\cite{footnoteN} Our goal is the description of the
flavor phase transition connected with generating the gap. We will
consider the dynamics both with and without an external magnetic field.

\section{Gap equation. Zero magnetic field}
\label{gapequat}

In this section we will describe the dynamics of the generation of a gap
connected with a quasiparticle-hole pairing provided by the interaction
(\ref{retard}) in the case of the zero external magnetic field. We will
begin by calculating the polarization function $\Pi(\omega,|\vec{k}|)$.
Actually, we will calculate (and use in the gap equation)  
$\Pi(0,|\vec{k}|)$, i.e., the polarization function in instantaneous
approximation. The reliability of this approximation will be discussed in
Sec.~\ref{inst}.

\subsection{Polarization function}
\label{polarization}

The one-loop polarization function at finite temperature and finite
chemical potential is given by the following integral representation
(see Appendix~\ref{AppA}):
\ba
\Pi(0, \vec{k}) &=& \frac{2Te^2N_f}{\varepsilon_{0} v_F^2}
\int_0^1 dx \Bigg[\ln \left(2\cosh\frac{R_{x}+\mu}{T}
\right)\nonumber\\
&-&\frac{\Delta_{T}^{2}(\mu)}{2TR_{x}}\tanh\frac{R_{x}+\mu}{2T}
+\left(\mu\to -\mu \right)
\Bigg],
\label{Pi-general}
\ea
where $R_{x}= \sqrt{v_F^2\vec{k}^2 x(1-x)+\Delta_{T}^{2}(\mu)}$, $\Delta_{T}
(\mu)$ is the fermion gap, and $T$ is temperature. Notice that the gap is
a dynamical quantity, determined from a gap equation (see
Sec.~\ref{dyngap} below), and therefore it can depend on both temperature
and chemical potential. Note
that throughout this paper we work in the vacuum in which the fermion gap
is positive.

At $\mu=0$ (zero density) and $T=0$, the polarization function 
becomes
\ba
\Pi(0, \vec{k}) &=& \frac{e^2N_f}{\varepsilon_{0} v_F^2} 
\Bigg(\Delta_{0} 
+ \frac{v_F^{2}\vec{k}^{2}-4\Delta_{0}^2}{2v_F|\vec{k}|}
\arctan\frac{v_F|\vec{k}|}{2\Delta_{0}}\Bigg).
\ea
At nonzero density and $T=0$, the function in  Eq.~(\ref{Pi-general}) 
reduces to
\ba
\Pi(0, \vec{k}) &=& \frac{2e^2N_f}{\varepsilon_{0} v_F^2}|\mu|, 
\mbox{ for } |\vec{k}|\leq k_{*},
\label{Pi_k<k0}\\
\Pi(0, \vec{k}) &=& \frac{2e^2N_f}{\varepsilon_{0} v_F^2}|\mu|\left[
1-\frac{\sqrt{\vec{k}^{2}-k_{*}^{2}}}{2|\vec{k}|}
+\frac{v_F^{2}\vec{k}^{2}-4\Delta_{0}^{2}(\mu)}{4\mu v_F|\vec{k}|}
\right. \nonumber \\ 
&\times& \left.
\arctan\frac{v_F\sqrt{\vec{k}^{2}-k_{*}^{2}}}{2\mu}\right],
\mbox{ for } |\vec{k}|> k_{*},
\label{Pi_k>k0}
\ea
where $k_{*}\equiv 2\sqrt{\mu^2-\Delta_{0}^2(\mu)}/v_F $ is proportional
to the square root of quasiparticle density at $T=0$, see
Eq.~(\ref{density-B0-T0}) below. As is easy to check, this polarization
function has a very strong dependence on momentum.  Indeed, while $\Pi(0,
\vec{k})$ remains constant for small momenta, $|\vec{k}|\leq k_{*}$, its
value drops considerably for $|\vec{k}|\gtrsim k_{*}$. In the case of a small
density of carriers, i.e., $n\sim k_{*}^{2} \ll (\Delta_{0}/v_F)^2$, this
momentum dependence is particularly strong. As is clear from Eq. (15), 
for small momenta, the polarization 
function $\Pi(0,\vec{k})$ is equal to the Debye mass $M_{D}$ and could be 
quite large. At the same time, the function $\Pi(0,\vec{k})$ 
at intermediate values of the momenta, $|\vec{k}| \sim
\sqrt{k_{*} \Delta_{0}(\mu)/v_F}$, is smaller than $M_{D}$ by 
about a factor of $\Delta_{0}(\mu)/\sqrt{\mu^2 -\Delta_{0}^2(\mu)}$, 
i.e., $\Pi(0,\vec{k})\sim 
M_{D}\sqrt{\mu^2-\Delta_{0}^2(\mu)}/\Delta_{0}(\mu)$. 
Finally, for $|\vec{k}|\gg \Delta_{0}(\mu)/v_F$, 
the polarization tensor approaches the following
asymptote:
\be
\Pi(0, \vec{k}) \simeq \frac{\pi e^2N_f}{4 \varepsilon_{0} v_F}
|\vec{k}|.
\label{polasympt}
\ee 
This observation is quite important for the proper analysis of the pairing
dynamics between electron and hole types of quasiparticles leading to a
possible dynamical generation of a gap. As we shall see below, it is in
fact the region of momenta $|\vec{k}| \gg \Delta_{0} (\mu)/v_F$ that
dominates in such a dynamics. This in particular implies that the one-loop
approximation with free gapless fermions (when both the gap and the wave
function renormalization are neglected) is a reliable approximation for
the polarization function in the gap equation, at least for large $N_{f}$.  
It could work reasonably well even for smaller values of $N_f$ of order $1$,
say, $2$ as in graphite.\cite{footnote}

\subsection{Dynamical gap at $T=0$ and $\mu=0$}
\label{dyngap}

Now, let us study a possibility of spontaneous generation of a dynamical
gap in the one-particle spectrum of quasiparticles. We begin by
considering the gap equation for the zero-density case  and 
zero-temperature case. Its explicit form
reads [see Eq.~(\ref{gap-A7}) in Appendix~\ref{AppA}]
\ba
\Delta_{p} = \lambda \int \frac{q dq \Delta_{q} 
{\cal K} \left(p, q \right)}
{\sqrt{q^{2}+(\Delta_{q}/v_{F})^{2}}},
\label{gap-eq}
\ea 
where the approximate expression for the kernel ${\cal K} (p,q)$ is 
given by 
\be
{\cal K} \left(p, q \right) = \frac{\theta(p-q)}{p}
+\frac{\theta(q-p)}{q},
\ee
and
\be
\lambda = \frac{e^{2}}{2(\varepsilon_{0}v_{F}+\pi e^{2}N_{f}/4)}.
\label{lambda}
\ee
It is well known,\cite{P,ANW,GGM} that the approximation with the one-loop
polarization function in the kernel of the gap equation (the so called
improved rainbow approximation) is reliable for large values of the number
of fermion flavors $N_{f}$. Here, however, we will also consider the
values of $N_{f}$ of order 1, say, 2 as in graphite. It is reasonable to
assume that this approximation works qualitatively (although apparently
not always quantitatively) even for these values of $N_{f}$, providing a
general insight into the nonperturbative dynamics of spontaneous
generation of a gap. The analysis done in $(2+1)$-dimensional QED supports
this assumption.\cite{QED_{2+1}}

In the most important region of momenta $|\vec{k}| \gg \Delta_{0}/v_F$
where the pairing dynamics dominates (see below), the only role of the
term $(\Delta_{q}/v_{F})^{2}$ in the denominator of the integrand on the
right hand side of Eq.~(\ref{gap-eq}) is to provide a cutoff in the
infrared region. Therefore one can drop this term, introducing instead the
explicit infrared cutoff $\Delta_{0}/v_{F}$ in the integral. This is the
essence of the so called bifurcation approximation. As a result, we arrive
at the following equation:
\ba
\Delta_{p} = \lambda  
\left(\int_{\Delta_{0}/v_{F}}^{p} \frac{dq }{p} \Delta_{q}
+\int_{p}^{\Lambda} \frac{dq }{q} \Delta_{q} \right).
\label{gap-eq-n0}
\ea
Here we also introduced a finite ultraviolet cutoff $\Lambda$. In a
condensed matter system, it could be taken of order $\pi/a$ where $a$ 
is a characteristic lattice size (for example, $a=2.46 \A$ for graphite). 
An alternative, equally good, estimate of $\Lambda$ is 
related to the size of the energy band $\Lambda = t/v_{F}$
where $t=2.4$ eV in the example of graphite.

The last integral equation is equivalent to the differential equation,
\be
p^{2}\Delta_{p}^{\prime\prime}+2 p \Delta_{p}^{\prime}
+\lambda \Delta_{p} =0,
\ee 
with the boundary conditions:
\ba
\left.
p^{2}\Delta_{p}^{\prime}\right|_{p=\Delta_{0}/v_{F}} &=& 0, 
\label{IR-bc}\\ 
\left.
\left(p \Delta_{p}^{\prime} + \Delta_{p} \right) 
\right|_{p=\Lambda} &=& 0.\label{UV-bc}
\ea
The solution compatible with the infrared boundary condition 
(\ref{IR-bc}) reads
\be
\Delta_{p} = \frac{\Delta_{0}^{3/2}}{\sin(\delta)\sqrt{pv_{F}}}
\sin\left(\frac{\sqrt{4\lambda-1}}{2}
\ln\frac{pv_{F}}{\Delta_{0}}+\delta\right),
\ee
where $\delta =\arctan\sqrt{4\lambda-1}$. Notice that $\Delta_{0}$
satisfies the relation $\Delta_{0}=\Delta_{p=\Delta_{0}/v_{F}}$.
The ultraviolet boundary condition (\ref{UV-bc}) imposes another 
restriction,
\be
\frac{\sqrt{4\lambda-1}}{2}
\ln\frac{v_{F}\Lambda}{\Delta_{0}}+2\delta=\pi.
\ee
As is clear from this equation, a meaningful solution for the dynamical
gap, satisfying the constraint $\Delta_{0}\ll \Lambda v_{F}$, exists only
for $\lambda > 1/4$. In the nearcritical region, i.e., when
$\sqrt{4\lambda -1}$ is small, the gap reads
\be
\Delta_{0} \simeq \Lambda v_{F} 
\exp\left(-\frac{2\pi}{\sqrt{4\lambda -1}}+4\right).
\label{gap-Del}
\ee
The condition $\lambda > 1/4$ gives the critical line in the plane
$(g,N_{f})$, where the dimensionless coupling constant is
$g \equiv e^{2}/{\varepsilon_{0} v_{F}}$:
\be
g_{cr} = \frac{4}{8-\pi N_{f}},
\label{crline}
\ee
which means that, in absence of an external magnetic field, a dynamical
gap is generated only if $g > g_{cr}$. In the example of graphite, the
number of ``flavors" is equal $2$.  Thus, the estimate of its critical
coupling gives $g_{cr} \approx 2.33$. We emphasize that it is just an
estimate obtained in the leading order in $1/N_{f}$ in instantaneous
approximation. For $N_{f} = 2$ as in graphite, both $1/N_{f}$ corrections
and improving the instantaneous approximation can certainly vary this
value (see a discussion in the end of Sec.~\ref{inst}).

If highly oriented pyrolytic graphite is a semimetal in absence of an
external magnetic field, it is clear that its effective coupling $g_{eff}$
(defined, for example, at the energy scale below which the Dirac type
effective action provides an appropriate description of the quasiparticle
dynamics) is smaller than $g_{cr}$. Indeed, if the interaction were
stronger than this, the ground state rearrangement (from a semimetal to an
insulator state), caused by the particle-hope pairing, could not be
prevented.

Let us now discuss the self-consistency of our assumption that the region
of momenta $|\vec{k}| \gg \Delta_{0}/v_F$ is mostly responsible for the
generation of a ``small" gap $\Delta_{0} \ll \Lambda v_{F}$ in the
nearcritical limit. The point is that in this regime the logarithm
$\ln(\Lambda v_{F}/\Delta_{0}) \sim {2\pi}/\sqrt{4\lambda -1}$ is large.
On the other hand, the behavior of the integrand on the right hand side of
Eq.~(\ref{gap-eq}) is smooth as $q \to 0$. The smooth behavior of the
integrand in the infrared region implies that the region $0 \leq q
\lesssim \Delta_{0}/v_{F}$ is too small to generate the large logarithm
$\ln(\Lambda v_{F}/\Delta_{0})$. This logarithm [and therefore the
essential singularity in expression (\ref{gap-Del})] is generated in the
large region $\Delta_{0}/v_{F} \ll q \ll \Lambda$. A variation of the
kernel in the infrared region can at most change the overall coefficient
in the expression for the gap.

At this point, we would like to mention that the dimensionless coupling
constant in the problem at hand is $g\equiv e^{2}/\varepsilon_{0}v_{F}$.  
In the gap equation, $g$ has to be considered as the bare coupling
constant and its value can be large. As was shown in
Ref.~\onlinecite{carbon}, in the absence of a dynamical gap, the
corresponding renormalization group (running) coupling runs
logarithmically to a trivial fixed point in infrared. In the presence of
the gap, such running should stop at the energy scale of order
$\Delta_{0}$. This means that the nonperturbative dynamics shifts the zero
infrared fixed point to a finite value.

\subsection{Dynamical gap at $T\neq 0$ or $\mu\neq 0$}

Up to now, we have considered the case with the zero density and zero
temperature. It is clear that the critical value of the coupling constant
should be larger than $g_{cr} \approx 2.33$ if a nonzero density (and/or
finite temperature) are taken into account. Indeed, with increasing the
charge density of carriers or the value of the temperature, the screening
effects get stronger and the quasiparticle interactions get weaker.  
In addition, the pairing between quasiparticles in the two adjacent bands
separated by the dynamical gap gradually becomes less efficient. The
latter could be clearly seen by comparing the energy gain from creating a
gap in the spectrum and the energy loss of pushing up the energy of all
the states in the band above the gap. Both effects work against the
formation of a gap. Thus, after reaching some critical value, the finite
density or temperature effects will be so strong that dynamical generation
of a gap will be impossible.

When the chemical potential is smaller than the gap, the dynamics of the
zero temperature model remains unchanged. Thus for all values of $\mu <
\Delta_{0} \equiv \Delta_{0}(\mu)|_{\mu=0}$, the exact solution for the
dynamical gap is the same. In our approximation, it is given by
Eq.~(\ref{gap-Del}). In order to consider the possibility of a nontrivial
solution satisfying the condition $\mu > \Delta_{0}(\mu)$, we consider an
approximate gap equation following from Eq.~(\ref{polasympt}) and 
Eqs.~(\ref{gap-A7}), (\ref{gap-eq-A9}) taken in the limit $T \to 0$:
\be
\Delta_{p} = \lambda
\left(\int_{\epsilon}^{p} \frac{dq }{p} \Delta_{q}
+\int_{p}^{\Lambda} \frac{dq }{q} \Delta_{q} \right),
\label{gap-eq-n}
\ee
where the infrared cutoff $\epsilon$ is given by a larger value of
$\Delta_{0}(\mu)/v_{F}$ or $\sqrt{\mu^{2}-\Delta_{0}^{2}(\mu)} /v_{F}$.  
By making use of the same method as before, we straightforwardly derive
two branches of the solution:
\be
\Delta_{0}(\mu) \simeq \Delta_0 ,
\label{gap-mu-1}
\ee
for $\mu < \sqrt{2} \Delta_0$ (here we took into account 
that $\Delta_{0}(\mu) =\Delta_0$ for $\mu < \Delta_0$), and 
\be
\Delta_{0}(\mu) \simeq \sqrt{\mu^2 -\Delta_0^2},
\label{gap-mu-2}
\ee 
valid for $\Delta_0 \leq |\mu| \leq \sqrt{2}\Delta_0$.  While the first
branch of the solution in Eq.~(\ref{gap-mu-1})  describes a gap that is
essentially unchanged with $\mu$, the value of the gap along the second
branch of the solution in Eq.~(\ref{gap-mu-2}) increases with the chemical
potential. For the values of the chemical potential in the range $\Delta_0
\leq \mu \leq \sqrt{2}\Delta_0$, both branches of the solution coexist.
The first branch corresponds to a locally stable solution (i.e., to a
local minimum of the effective potential), while the other one --- to an
unstable solution (i.e., to a local maximum of the effective potential).
In addition, there is always a trivial solution which corresponds to an
extremum of the effective potential at the origin. When both nontrivial
solutions (\ref{gap-mu-1}) and (\ref{gap-mu-2}) coexist, the extremum at
the origin should be a minimum. This follows from a simple consideration
of the topology of the effective potential. 
\begin{figure}
\epsfxsize=8.0cm
\epsffile[90 -5 580 290]{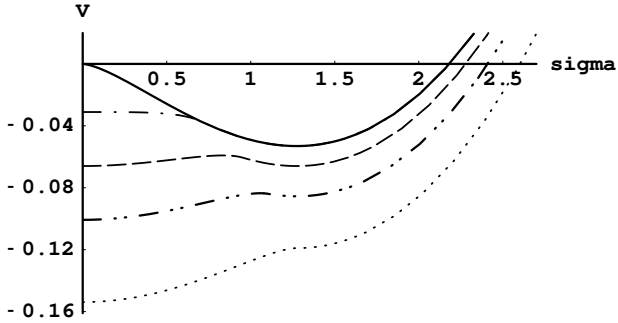}
\caption{The effective potential of the composite field $\sigma$ for a few
different values of the chemical potential: $\mu=0$ (solid line),
$\mu=\Delta_{0}$ (dash-dotted line), $\mu=\mu_{c}\approx 1.19\Delta_{0}$
(dashed line), $\mu=1.3\Delta_{0}$ (dash-dot-dotted line), 
and $\mu=\sqrt{2}\Delta_{0}$ (dotted line). In calculation,
we used $\Delta_0=1$. The values of the potential are given in units of
$N_{f}\Delta_{0}^3/v_{F}^{2}$, and values of the composite field $\sigma$
--- in units of $N_{f}(\Delta_{0}/v_{F})^{3/2} \sqrt{\Lambda}$.}
\label{fig:eff-pot}
\end{figure}
Of course, the analysis of the gap equation alone would not be sufficient
to prove most of the above statements. To support them, we derived the
effective potential $V(\sigma)$ as a function of the composite field
$\sigma= -\langle \bar{\psi}\psi\rangle$ in Appendix~\ref{AppB}. This
potential is graphically shown in Fig.~\ref{fig:eff-pot} for a few
different values of the chemical potential. As is clear from the figure,
we have a typical realization of the first order phase transition.

Our analysis of the effective potential also allows to determine the
critical value of the chemical potential:
\ba
\mu_{c} &\simeq & \frac{\Delta_0}{(2-\sqrt{2})^{1/3}} \nonumber \\
&\simeq &\frac{\Lambda v_{F}}{(2-\sqrt{2})^{1/3} }
\exp\left(-\frac{2\pi}{\sqrt{4\lambda -1}}+4\right).
\ea
When the chemical potential increases from $\mu=\mu_{c}-0$ to
$\mu=\mu_{c}+0$, the value of the gap drops from $\Delta\simeq\Delta_0$
down to $\Delta=0$.

Similarly, at $\mu=0$, we could derive the value of the critical 
temperature. It also appears to be of the same order as $\Delta_{0}$:
\be
T_{c} \simeq \frac{\Delta_{0}}{2}
\simeq \frac{\Lambda v_{F}}{2}
\exp\left(-\frac{2\pi}{\sqrt{4\lambda -1}}+4\right).
\ee
Unlike the case with the chemical potential, the phase transition in
temperature is of the second order. This follows both from the
existence of a single-branch solution to the gap equation and 
from a direct study of the effective potential.

\subsection{Reduced QED$_{3+1}$ vs. conventional QED$_{2+1}$}
\label{vs}

Before concluding this section, it is instructive to compare the reduced
dynamics with the ``conventional"
QED$_{2+1}$. The gap equations in these two models are similar, but the
interaction potentials are slightly different. Instead of expression
(\ref{retard}), one has\cite{P,ANW}
\be
U_{3d}(t,\vec{r})= \frac{e^{2}_{3}}{\varepsilon_{0}}
\int\frac{d\omega}{2\pi} \int \frac{d^{2}\vec{k}}{(2\pi)^{2}}
\frac{\exp(-i\omega t+i\vec{k}\cdot\vec{r})}
{\vec{k}^{2}+\Pi(\omega,|\vec{k}|)},
\label{3Dretard}
\ee
where $e_{3}$ is the (dimensionful) coupling constant in QED$_{2+1}$ 
and, in the relevant region of momenta $\Delta/v_{F} \ll  |\vec{k}| \ll
N_{f}e_{3}^{2}/\varepsilon_{0}v_{F}$, the polarization function here is
essentially the same as in Eq.~(\ref{Pi-general}), except that the
dimensionless coupling constant $e^{2}/\varepsilon_{0}v_{F}$ is replaced by
the dimensionful $e_{3}^{2}/\varepsilon_{0}v_{F}$. Comparing expressions
(\ref{3Dretard}) and (\ref{retard}), one can see that the only difference
between them is in the appearance of the term $\vec{k}^{2}$, instead
$|\vec{k}|$, in the denominator of the former.  This point makes quite a
difference. On the one hand, it provides a dynamical ultraviolet cutoff
$\sim N_{f} e^2_{3}/\varepsilon_{0}v_{F}$ 
in the gap equation and, on the other hand, since this term
is suppressed in the region 
$|\vec{k}| \ll N_{f} e_{3}^{2}/\varepsilon_{0}v_{F}$, 
it is irrelevant for generating the gap. This implies reducing
screening of Coulomb like interactions in QED$_{2+1}$ as compared to the
reduced dynamics.  Let us consider this point in more detail. It is easy
to find that the dynamical gap in QED$_{2+1}$ is
\be
\Delta_{3d} \simeq \frac{N_{f} e_{3}^{2}}{\varepsilon_{0}} 
\exp\left(-\frac{2\pi}{\sqrt{4\lambda_{3}-1}}+4\right),
\ee
where $\lambda_{3}=2/\pi N_{f}$. Since this solution exists when
$\lambda_{3}>1/4$, it implies that the critical value of $N_{f}$ is equal
to $8/\pi\approx 2.55$. The same critical value for $N_{f}$ was
obtained in Ref.~\onlinecite{Khvesh2}.

Now, notice that the parameter $\lambda_{3}$ coincides with $\lambda$ in
Eq.~(\ref{lambda}) only in the limit $e^{2}\to \infty$. Thus, the reduced
dynamics becomes equivalent to QED$_{2+1}$ dynamics only in the maximally
strong coupling limit, with $e^{2}\to \infty$. Therefore, we conclude that
there are important similarities and important differences between the
dynamics in QED$_{2+1}$ and reduced QED$_{3+1}$. Both
dynamics are intimately connected with long range Coulomb like
interactions. On the other hand, since QED$_{2+1}$ is superrenormalizable
(and therefore asymptotically free) theory, its nonperturbative
interactions are dynamically cut off at the scale 
$\sim N_{f} e_{3}^{2}/\varepsilon_{0}v_{F}$ in
ultraviolet. Also, its dynamics is more efficient in generating a
dynamical gap. Indeed, it corresponds to the dynamics in the reduced
QED$_{3+1}$ when the coupling constant $e$ of the latter goes to infinity.
This feature has been already established in relativistic reduced
QED$_{3+1}$,\cite{GGM} with $v_{F}=c$.

\subsection{Beyond instantaneous approximation}
\label{inst}

Let us now turn to the discussion of the reliability of the instantaneous
approximation for the gap equation. In this approximation, the frequency
dependence in the photon propagator is neglected. While it is certainly
justified for its free (kinetic) part, it is not immediately clear how
good it is for the polarization function.

Keeping the frequency dependence, the gap equation at zero $T$ and $\mu$
takes the following form in Euclidean space [compare with Eq.~(\ref{B7})
in Appendix~\ref{gapB}]:
\ba
\Delta(\Omega,p) &=& \frac{e^2}{\varepsilon_{0}} \int
\frac{d\omega d^2k}{(2\pi)^2} \frac{\Delta(\omega ,k)}
{\omega^2+v_F^2 |\vec{k}|^2+\Delta(\omega ,k)^2}
\nonumber \\
&\times& \frac{1}{|\vec{p}-\vec{k}|
+\Pi(\Omega-\omega,\vec{p}-\vec{k})},
\label{noninst}
\ea
where the approximate polarization function is 
\be
\Pi(\omega,\vec{k}) =\frac{\pi e^2 N_f |\vec{k}|^2}
{4\varepsilon_{0}  \sqrt{\omega^2+v_{F}^{2}|\vec{k}|^2}}.
\label{polomega}
\ee
In the instantaneous approximation, we used $\omega^2 \to 0$ in the
polarization function. Thus, the strength of the interaction was somewhat
underestimated in such an approximation.  This in turn implies that
$g_{cr}$ and $1/N_{f}^{cr}$ are smaller than their values
obtained in the instantaneous approximation, i.e.,
$g_{cr}< 4/(8-\pi N_{f})$ and $N_{f}^{cr}>8/\pi\approx 2.55$
in the leading order in $1/N_{f}$.

One should remember however that $1/N_{f}$ corrections could
be relevant for the values of $N_{f}$ of order 1. Using the argument in
Ref.~25, one may expect the variations up to 50 percent
in the value of $N_{f}^{cr}$. Therefore the value $N_{f}^{cr} = 2.55$,
obtained here in leading order in $1/N_{f}$, should be considered just as
a useful estimate.

\section{Gap equation. Nonzero magnetic field}
\label{sec:mag-field}

The main goal of this paper is a description of the 
magnetic-field-driven metal-insulator phase transition in 
planar systems. Having
developed a general formalism in the preceding sections, here we will take
into account the effect of an external constant magnetic field on the
dynamics of a
spontaneous generation of a gap.\cite{newfootnote} 
The general observation of
Refs.~\onlinecite{1,2} states that in the presence of an external 
magnetic
field, there is the generation of a dynamical gap connected with
electron-hole pairing even for an arbitrary weak attraction between
electrons and holes. Therefore in this case the gap will appear even if
the bare coupling constant $g$, introduced in the previous section, is
subcritical (in the case of a supercritical $g$, the magnetic field would
enhance the already existent gap). This phenomenon is known as ``magnetic
catalysis". The origin of this effect is connected with the dynamics of
the LLL: its dynamics is effectively $0 + 1$ ($1 + 1$) dimensional in $2 +
1$ ($3 + 1$) dimensions and this makes the electron-hole pairing
inevitable.

Actually, this formulation is correct only in the case of zero temperature
and zero charge density.  In the presence of temperature $T$ and/or charge
density $n$, there is a critical value of the magnetic field,
$B_{c}(T,n)$, defining a threshold for this effect: $B$ has to be larger
than $B_c(T,n)$.\cite{1} While the dependence of $B_c(T,n)$ on $T$ is
model dependent, the value of $B_c(0,n)$ is universal for all values of $g
\leq g_c$: it is given by $|e|B_c = 2\pi c n/N_{f}$ and corresponds to the
filling of the lowest Landau level.\cite{1} The physics of this result is
quite clear: when the LLL is filled up, the LLL electrons are blocked and
excluded from the pairing dynamics.  In other words, in this case we loose
the catalyst and, therefore, the effect itself. As we will see in
Sec.~\ref{HOPG}, this point can be crucial for explaining the presence of
an offset field $B_c$ observed in the recent experiments in highly
oriented pyrolytic graphite.\cite{Exp1,SSCom115,Exp2}

\subsection{Dynamical gap at $T=0$ and $\mu=0$}
\label{subsec:T0mu0}

In a constant external magnetic field $B$, only the free part of the
quasiparticle action is modified. In particular, the spatial derivatives
in Eq.~(\ref{L-free}) are replaced by the corresponding covariant
derivatives:
\ba
\partial_{x} &\to& \partial_{x} + i \frac{e}{c} A^{ext}_{x}(\vec{r}), \\  
\partial_{y} &\to& \partial_{y} + i \frac{e}{c} A^{ext}_{y}(\vec{r}),
\ea 
where $A^{ext}_{x}(\vec{r})=-By/2$ and $A^{ext}_{y}(\vec{r})= Bx/2$.  In
this case, the propagator of quasiparticles takes the following general
form:\cite{2,4}
\be
G(t-t^{\prime},\vec{r},\vec{r}^{\prime}) = 
\exp\left[-i\frac{e}{c}\vec{r}\cdot \vec{A}^{ext}(\vec{r}^{\prime}) \right]
\tilde{G} (t-t^{\prime},\vec{r}-\vec{r}^{\prime}), 
\label{G-phase}
\ee
Notice that while we used the symbol $S$ for the fermion propagator in the
case without magnetic field, we use the symbol $G$ for the fermion
propagator in a magnetic field.  Let us begin by considering the
propagator of free quasiparticles in a magnetic field,
$G_{0}(t-t^{\prime},\vec{r},\vec{r}^{\prime})$.  For our purposes, it will
be useful to introduce the bare gap $\Delta_{b}$ for these free
quasiparticles.  The translation invariant part of such a propagator,
$\tilde{G_{0}}(t-t^{\prime},\vec{r}-\vec{r}^{\prime})$, reads\cite{1}
\ba
\tilde{G_{0}} (t, \vec{r}) &=& \int \frac{dt}{2\pi} 
\frac{d^{2} \vec{k}}{(2\pi)^{2}} 
\exp(-i\omega t +i \vec{k}\cdot\vec{r} ) \tilde{G_{0}} (\omega,\vec{k}), 
\label{G-Fourier}\\
\tilde{G_{0}} (\omega,\vec{k}) &=& 
2 i \exp\left(-\frac{c |\vec{k}|^{2}}{|eB|}\right)
\nonumber \\
&\times&
\sum_{n=0}^{\infty} \frac{(-1)^{n} \left[(\omega\gamma^{0} +\Delta_{b})
f_{1}({\vec{k}}) + f_{2}({\vec{k}})\right]} 
{\omega^{2}-\Delta_{b}^{2}-2nv_{F}^{2}|eB|/c}.
\label{G-LLs}
\ea
In this last equation, we used the shorthand notation:
\ba
f_{1}({\vec{k}}) &=& {\cal P}_{-}L_{n}\left(
\frac{2c{\vec{k}}^{2}}{|eB|}\right)
-{\cal P}_{+}L_{n-1}\left(\frac{2c{\vec{k}}^{2}}{|eB|}\right) ,
\label{expr:f1} \\
f_{2}({\vec{k}}) &=& 2 v_{F} {\vec{k}}{\vec{\gamma}}L_{n-1}^{1}
\left(\frac{2c{\vec{k}}^{2}}{|eB|} \right).
\label{expr:f2}
\ea
with the following spin projection operators:
\be
{\cal P}_{\pm} = \frac{1\pm i\gamma^{1}\gamma^{2}}{2}.
\ee
Also, $L^{\alpha}_{n}(z)$ are the generalized Laguerre polynomials. By
definition, $L_{n}(z)\equiv L^{0}_{n}(z)$ and $L^{\alpha}_{-1}(z)\equiv
0$.

Let us now turn to the interactions in the presence of the magnetic
field. In this case the polarization effects could also be taken into
account,\cite{Shpagin} and the modified interaction is:
\ba
U(t,\vec{r}) &=& \delta(t)
\frac{e^2}{\varepsilon_{0}} \int \frac{d^{2}\vec{k}}{2\pi} 
\frac{\exp(i\vec{k}\cdot\vec{r})}{|\vec{k}|(1+a |\vec{k}|)} \nonumber\\
&=& \frac{e^{2}\pi \delta(t)}{2\varepsilon_{0}a}
\left[H_{0}\left(\frac{|\vec{r}|}{a}\right)
-N_{0}\left(\frac{|\vec{r}|}{a}\right)\right],
\ea
where 
\be
a= 2\pi\nu_{0} \frac{e^2 N_{f}}{\varepsilon_{0}v_{F}}
\sqrt{\frac{c}{|eB|}},
\ee
and the constant $\nu_{0}$ is given by
\ba
\nu_{0} &\equiv& \frac{1}{4\pi\sqrt{\pi}} \int_{0}^{\infty}
\frac{dz}{\sqrt{z}} \left(\frac{\coth(z)}{z}-\frac{1}{\sinh^{2}(z)}\right)
\nonumber\\
&=& -\frac{3\zeta(-0.5)}{\sqrt{2}\pi}\approx 0.14. 
\ea
Regarding the new notation, $\zeta(z)$ is the Riemann zeta function,
$H_{0}(z)$ is the Struve function, and $N_{0}(z)$ is the Bessel function
of the second kind. It is noticeable that the instantaneous approximation
for the polarization function is justified in this case: the frequency
dependence is suppressed by factors of order $\omega/\sqrt{v_{F}|eB}|$
(which are small in the case of the LLL dominance (see below)). This can
be shown directly from the expression for the polarization function in
Ref.~\onlinecite{Shpagin}.

Now, the gap equation for the quasiparticle propagator reads
\ba
\tilde{G}(t,\vec{r}) &=& \tilde{G}_{0}(t,\vec{r}) 
-i \int dt^{\prime} d^{2} \vec{r}^{\prime} 
\int dt^{\prime\prime} d^{2} \vec{r}^{\prime\prime} \nonumber\\
&\times& \exp\left[-i\vec{r}\cdot \vec{A}(\vec{r}^{\prime})
-i\vec{r^{\prime}}\cdot \vec{A}(\vec{r}^{\prime\prime})\right] 
\nonumber\\
&\times& \tilde{G}_{0}(t-t^{\prime},\vec{r}-\vec{r}^{\prime}) 
\gamma^{0} \tilde{G}(t^{\prime}-t^{\prime\prime},
\vec{r}^{\prime}-\vec{r}^{\prime\prime}) \nonumber\\
&\times&\gamma^{0} 
\tilde{G}(t^{\prime\prime},\vec{r}^{\prime\prime})
U(\vec{r}^{\prime}-\vec{r}^{\prime\prime})
\delta(t^{\prime}-t^{\prime\prime}).
\label{SD-eq}
\ea
The structure of this equation is essentially the same as in the
relativistic model of Refs.~\onlinecite{2,4}. Here, however, we neglect
the retardation effects in the interaction potential. 

As was pointed out in Ref.~\onlinecite{1}, in the case of a subcritical
coupling constant $g \leq g_{c}$, one should distinguish two different
dynamical regimes.  The first regime corresponds to the situation with a
weak coupling $g$, when it is outside the scaling region near the critical
value $g_c$. In this case the LLL dominates and the value of the dynamical
gap $\Delta_0$ is much less than the gap $\sqrt{2v_{F}^2|eB|/c}$ between
the Landau levels. The latter guarantees that the higher Landau levels
decouple from the pairing dynamics and the LLL dominates indeed.

The second, strong coupling, regime is that with a near-critical, although
subcritical, value of $g$. In that case, all Landau levels are relevant
for the pairing dynamics and the value of the dynamical gap $\Delta_0$ is
of order of the Landau gap $\sqrt{2v_{F}^2|eB|/c}$.

Let us begin by considering the first regime. Then, the low energy
dynamics is dominated by the LLL, and the quasiparticle propagator could
be approximated as follows:
\be
\tilde{G}(t,\vec{r}) = \frac{i|eB|}{4\pi c} 
\exp\left(-\frac{|\vec{r}|^{2} |eB|}{4c}\right)
g(t)(1-i\gamma^{1}\gamma^{2}),
\label{LLL-app}
\ee 
where $g(t)$ is unknown matrix-valued function which should be determined
by solving the Schwinger-Dyson (gap) equation. By substituting the ansatz
(\ref{LLL-app}) into Eq.~(\ref{SD-eq}), we derive the following equation
for the Fourier transform of $g(t)$:
\ba
g^{-1}(\omega) &=& g_{0}^{-1}(\omega) 
- i e^{2} \int \frac{d \omega^{\prime}}{2\pi}
\gamma^{0} g(\omega-\omega^{\prime}) \gamma^{0}\nonumber\\
&\times& \int \frac{d^{2} \vec{k}}{(2\pi)^{2}} 
\exp\left[-\frac{c|\vec{k}|^{2}}{2|eB|}\right] 
U(\vec{k}).
\label{eq-omega}
\ea
The value of the bare gap is now zero in the free propagator
$g_{0}(\omega)$. And the general structure of the function $g(\omega)$ is
suggested by the first (LLL) term in the bare propagator (\ref{G-LLs}),
where now the bare gap $\Delta_{b}$ should be replaced by the dynamical
gap function $\Delta_{\omega}$ and the wave function renormalization
$A_{\omega}$ should be introduced. Thus, we have
\be
g(\omega) = \frac{A_{\omega}\gamma^{0}\omega +\Delta_{\omega}}
{A_{\omega}^{2}\omega^{2}-\Delta_{\omega}^{2}}.
\ee
One could see that the integral on the right hand side of
Eq.~(\ref{eq-omega}) is independent of $\omega$. This implies that
$A_{\omega}=1$ and the gap $\Delta_{\omega}$ is independent of $\omega$.
By taking this into account, we straightforwardly derive the solution:
\be
\Delta_{0} = \frac{g}{\sqrt{2}} \sqrt{\frac{v_{F}^{2}|eB|}{c}}
\int_{0}^{\infty} 
\frac{dk \exp(-k^2)}{1+k \chi_{0}} ,
\label{gap-T0}
\ee
where 
\be
\chi_{0} = 2\sqrt{2}\pi\nu_{0} g N_{f}.
\ee
In two limiting cases, $\chi_{0} \ll 1$ and $\chi_{0}\gg 1$,
we get the following asymptotes:
\be
\Delta_{0} \simeq  \frac{g\sqrt{\pi}}{2\sqrt{2}} 
\sqrt{\frac{v_{F}^{2}|eB|}{c}} \left(1-\frac{\chi_{0}}{\sqrt{\pi}} +
\frac{\chi_{0}^{2}}{2}+\cdots\right),
\label{weak}
\ee
(for weak coupling and small $N_{f}$) and
\be 
\Delta_{0} \simeq \frac{g}{\sqrt{2}} 
\sqrt{\frac{v_{F}^{2}|eB|}{c}} \frac{\ln\chi_{0}}{\chi_{0}}
\equiv \frac{v_{F}}{4\pi\nu_{0}N_{f}}\sqrt{\frac{|eB|}{c}}\ln\chi_{0},
\label{largeN_f}
\ee
(for large $N_{f}$). In accordance with the general 
conclusion of Refs.~\onlinecite{1,2}, in a magnetic field the gap is
generated for any nonzero coupling constant $g=e^2/\varepsilon_{0}v_F$.

One can see that for a sufficiently small $g=e^2/\varepsilon_{0}v_F$ in
expression (\ref{weak}) and for a sufficiently large $N_f$ in
(\ref{largeN_f}), the LLL approximation is selfconsistent indeed: in both
cases, the gap $\Delta_0$ can be made much less than the Landau gap
(scale) $L \equiv \sqrt{2v_{F}^2|eB|/c}$.  We emphasize that the second
solution (\ref{largeN_f}), obtained also in Ref.~\onlinecite{Khvesh},
corresponds to the regime with a large $N_{f}$ and {\it not} to the strong
coupling regime with a large $g$ and $N_f$ of order one. Indeed, taking
$g$ to be large enough in expression (\ref{largeN_f}), one gets the gap
$\Delta_0$ exceeding the Landau scale $L$, i.e., for large $g$ the
self-consistency of the LLL dominance approximation is lost. We will
discuss the strong coupling regime below.

What is the energy scale the coupling constant $g$ relates to in this
problem? It is the Landau scale $L$. The argument supporting this goes as
follows. There are two, dynamically very different, scale regions in this
problem. One is the region with the energy scale above the Landau scale
$L$ and below the ultraviolet cutoff $\Lambda$, defined by the lattice
size. In that region, the dynamics is essentially the same as in the
theory without magnetic field. In particular, the running coupling
decreases logarithmically with the energy scale there.\cite{carbon}
Another is the region below the Landau scale. In that region, the magnetic
field dramatically changes the dynamics, in particular, the behavior of
the running coupling constant.  As the analysis of this section shows,
because of the magnetic field, the pairing dynamics (in the particle-hole
channel) is dominated by the infrared region where $\omega \lesssim
\Delta_{0}$. Therefore, the scale region above the Landau scale $L$
completely decouples from the pairing dynamics in this case. This
manifests itself in expression (\ref{gap-T0}) for the gap: the only
relevant scale is the Landau scale $L$ there. Since the effect of the
running of the coupling is taken into account by the polarization function
in the gap equation, we conclude that the coupling $g$ indeed relates to
the Landau scale in this problem. Notice that it can be somewhat smaller
than the bare coupling constant $g(t)$ related to the scale
$t$. Taking $t = 2.4$ eV in graphite (the width of its energy
band) and using the equation for the running coupling from
Ref.~7, one obtains that it is smaller by the factors 1.2
and 1.4 than $g(t)$ for the values of the magnetic field $B = 10$ T
and $B = 0.1$ T, respectively.

Now let us turn to the second, strong coupling, dynamical regime. In
reduced QED, the gap equation in this regime includes the contributions of
all the Landau levels and becomes very formidable. Still, one can estimate
the value of the gap: since there are no small parameters in this regime
for moderate values of $N_f$, the gap should be of the order of the Landau
scale $L$. This conclusion is supported by studying this regime in a
simpler model, $(2+1)$-dimensional Nambu-Jona-Lasinio model.\cite{1} In that
case, in the critical regime, the gap equals $\Delta_{0} \simeq 0.32 L$,
where the Landau scale in that relativistic model, with $v_{F}=c$, is $L =
\sqrt{2 c |eB|}$. As we will see in Sec.~\ref{HOPG}, the critical
dynamical regime can be relevant for the magnetic-field-driven phase
transition in highly oriented pyrolytic graphite.\cite{Exp1,SSCom115,Exp2}

\subsection{Dynamical gap at $T\neq 0$ or $\mu\neq 0$}
\label{neq} 

By making use of the Matsubara formalism, it is easy to generalize 
the gap equation for the case of a finite temperature and a nonzero 
chemical potential. Without going into all details, let us write
down the final equation, 
\ba
\Delta_{T}(\mu) &=&
\frac{e^{2}}{2\sqrt{2}\varepsilon_{0}}\sqrt{\frac{|eB|}{c}}
\int_0^{\infty} \frac{dk \exp(-k^2)}{1+k\chi_0} \nonumber \\
&\times& 
\frac{\sinh\frac{\Delta_{T}(\mu)}{T}}{\cosh\frac{\Delta_{T}(\mu)}{T}
+\cosh\frac{\mu}{T}}.
\ea
In the LLL dominance approximation, the expression for charge density of 
carriers in terms of the chemical potential is
\ba
n&=&\frac{N_{f}|eB|}{2\pi c}
\frac{\sinh\frac{\mu}{T}}{\cosh\frac{\Delta_{T}(\mu)}{T}
+\cosh\frac{\mu}{T}}.
\label{density}
\ea
We assume that, in the model at hand, the charge density 
of carries (i.e., $n=n_{el}-n_{h}$) is a fixed constant. Then, 
the expression for the chemical potential reads
\ba
\sinh\frac{\mu}{T} &=&\frac{\nu_{B}}{1-\nu_{B}^{2}}\Bigg(
\cosh\frac{\Delta_{T}(\mu)}{T} \nonumber\\
&+&\sqrt{1+\nu_{B}^{2}\sinh^{2}\frac{\Delta_{T}(\mu)}{T}}\Bigg), \\
\cosh\frac{\mu}{T} &=&\frac{1}{1-\nu_{B}^{2}}\Bigg(
\sqrt{1+\nu_{B}^{2}\sinh^{2}\frac{\Delta_{T}(\mu)}{T}}
\nonumber\\
&+&\nu_{B}^{2}\cosh\frac{\Delta_{T}(\mu)}{T}\Bigg),
\ea 
where 
\be
\nu_{B} = \frac{2\pi c n}{N_{f}|eB|} \equiv \frac{B_{c}}{B}
\ee
is the filling factor. 

By making use of the expression for the chemical potential in terms of the
filling factor $\nu_{B}$, we rewrite the gap equation in the following
convenient form:
\ba
\Delta_{T}(\nu_{B}) &=& 
\frac{e^{2}}{2\sqrt{2}\varepsilon_{0}}\sqrt{\frac{|eB|}{c}}
\int_0^{\infty} \frac{dk \exp(-k^2)}{1+k\chi_0} \nonumber \\
&\times& 
\frac{(1-\nu_{B}^{2})\sinh\frac{\Delta_{T}(\nu_{B})}
{T}}{\cosh\frac{\Delta_{T}(\nu_{B})}{T}
+\sqrt{1+\nu_{B}^{2}\sinh^{2}
\frac{\Delta_{T}(\nu_{B})}{T}}}.
\label{gapT}
\ea
Let us first consider the case of zero temperature. Then the
gap equation takes a very simple form:
\ba
\Delta_{0}(\nu_{B}) &=& \frac{1-\nu_{B}}{2} \Delta_{0} 
\nonumber \\
&\equiv &\frac{e^{2}(1-\nu_{B})}{2\sqrt{2}\varepsilon_{0}}
\sqrt{\frac{|eB|}{c}} \int_0^{\infty}
\frac{dk \exp(-k^2)}{1+k\chi_0},
\label{nu_c}
\ea
where $\Delta_{0} \equiv \Delta_{0}(0)$ is the value of the gap at the
zero filling factor. Since we choose the vacuum in which the gap is
positive (see Sec.~\ref{polarization}), this equation implies that for
$\nu_{B} \geq 1$, there is no solution with a nonzero gap, i.e. the
symmetry is restored. The condition $\nu_{B} = 1$ determines the critical
density, $n_{c} = N_{f}|eB|/2\pi c$. The density $n_c$ corresponds to the
filling of the LLL and, as was discussed at the beginning of this section,
this value is universal for all subcritical values of the coupling
constant $g$. Reversing the roles of $n$ and $B$, one can say that, for a
fixed value of the density $n$, the critical value of the magnetic field
is $|eB_c| = 2\pi c n/N_f$: a dynamical gap occurs only for magnetic
fields $B$ larger than $B_c$.

The critical temperature is determined from Eq.~(\ref{gapT}) 
with $\Delta_{T}(\nu_{B}) = 0$:
\be
T_{c} = \frac{e^{2}(1-\nu_{B}^{2})}{4\sqrt{2}\varepsilon_{0}}
\sqrt{\frac{|eB|}{c}} \int_0^{\infty} 
\frac{dk \exp(-k^2)}{1+k\chi_0}.
\label{T_c}
\ee
At a fixed density $n$, this equation implies that, as it should be, the
critical temperature is zero for magnetic fields weaker than the critical
value $B_{c}$ determined above. For magnetic fields $B$ stronger than
$B_c$, $T_c$ grows with $B$ (see Fig.~\ref{Tcrit} in Sec.~\ref{HOPG}). As
we will see in Sec.~\ref{HOPG}, these results can be important for
explaining experimental data in highly oriented pyrolytic
graphite.\cite{Exp1,SSCom115,Exp2}

Though here we considered only the dynamical regime with the LLL
dominance, it is reasonable to assume that the qualitative picture will
remain the same also in the case of the scaling dynamical regime, with the
near-critical coupling constant $g$. This is in particular supported by
the fact of the universality of the critical value $B_c$.

Before concluding this section, let us mention that the
the gap equation could also be rewritten in the following form:
\be
\Delta_{T}(\nu_{B}) = \frac{2 T_{c}
\sinh\frac{\Delta_{T}(\nu_{B}) }{T}}
{\cosh\frac{\Delta_{T}(\nu_{B})}{T}+\sqrt{1+\nu_{B}^{2}\sinh^{2}
\frac{\Delta_{T}(\nu_{B})}{T}}},
\label{gap-Tc}
\ee
where the relation (\ref{T_c}) was taken into account. The last form of
the gap equation will be the most convenient for using in numerical
calculations of conductivity; see Sec.~\ref{sec:cond-mag}.
 
\section{Conductivity and Resistivity}
\label{condres}

Conductivity and resistivity are major players in expe\-rimental detecting
the magnetic field driven semimetal-insulator phase transition in
graphite.\cite{Exp1,SSCom115,Exp2} In this section, we will calculate them
in reduced QED$_{3+1}$ using the results obtained in the
previous sections. We will consider both cases of zero and nonzero external
magnetic fields. While the former case is interesting in itself, it will
also serve us as an important reference point for the latter. Tne main
conclusion of this section is that there is a clear signature of the phase
transition seen in the behavior of the conductivity (resistivity) as a
function of temperature. More precisely, we find that
\begin{enumerate}
\item if the phase transition is of the first order, there is a
    discontinuity in the conductivity (resistivity) at a 
    critical temperature $T_{c}$;

\item if the phase transition is of the second order and the scaling 
    properties are correctly described by the mean-field approximation,
    the conductivity (resistivity) exhibits a kink behavior at the
    critical temperature;

\item at last, if the phase transition is a non-mean-field
    second order one,
    the conductivity (resistivity) is a smooth function at the
    critical temperature $T_c$, while a singularity  
    occurs in its higher derivatives at $T=T_{c}$.   
\end{enumerate}
Besides, our calculations show that in this particular model, the flavor
phase transition, restoring the flavor symmetry $U(2N_F)$, does not look as
a semimetal-insulator phase transition {\it if} there is no external
magnetic field. On the other hand, in the presence of a magnetic field, in
many cases it does look as a semimetal-insulator phase transition.

\subsection{No Magnetic Field}
\label{nomag}

In calculation of transport coefficients, it is very useful to utilize the
spectral representation of the quasiparticle Green function. The latter is
defined as follows:\cite{footnote1}
\be
S(i\omega _{n},{\vec{k}})=i \int\limits_{-\infty }^{\infty }
\frac{d\omega A(\omega ,{\vec{k}})}
{i\omega _{n}-\mu-\omega }.  
\label{spectr_repr}
\ee
In the reduced planar model described in Sec.~\ref{sec:model}, 
we derive
\be
A(\omega ,{\vec{k}})=\frac{\Gamma}{2\pi E}
\left[\frac{\gamma^0E-{\vec{k}}\vec{\gamma}+\Delta}
{(\omega-E)^2+\Gamma^2}+
\frac{\gamma^0E+{\vec{k}}\vec{\gamma}-\Delta}{(\omega+E)^2+\Gamma^2}
\right],
\ee
where $E=\sqrt{v_F^2{\vec{k}}^2+\Delta^2}$ and, throughout this section,
we use the symbol $\Delta$ for the gap, i.e., $\Delta \equiv
\Delta_{T}(\mu)$ and $\Delta \equiv \Delta_{T}(\nu_{B})$ for the cases
with no and with magnetic field, respectively.  Notice that we introduced
a phenomenological width parameter $\Gamma$ without which the calculation
of conductivities would be meaningless.  A finite width parameter appears
as a result of interactions, and scattering on impurities, in particular.
In general, the width $\Gamma$ is defined through the fermion self-energy
as $\Gamma(\omega)= -\mbox{Im} \Sigma^R(\omega)$. Thus, it is a frequency
(as well as temperature and magnetic field) dependent quantity. Like the
dynamical gap itself, it should be self-consistently determined from the
Schwinger-Dyson equations. At low temperatures, usually it could be
modeled by a constant phenomenological parameter. Therefore, instead of
considering an additional Schwinger-Dyson type equation, we choose a
constant parameter $\Gamma$ and view this as yet another approximation.

In terms of the spectral function, the charge density of carriers 
reads
\be
n = \frac{1}{2}\int\frac{d^2k}{(2\pi)^2}
\int\limits_{-\infty}^\infty d\omega
\left(\tanh\frac{\omega+\mu}{2T} -1\right)\mbox{tr}
\left[\gamma^0 A(\omega,{\vec{k}})\right].
\ee
The conductivity tensor is defined as follows:
\be
\sigma_{ij} = \lim_{\Omega\to 0} \frac{\mbox{Im}
\Pi^R_{ij}(\Omega+i\epsilon)}{\Omega},
\ee
where $\Pi^R_{ij}(\Omega)$ is the retarded current-current correlation 
function which is also given in terms of the spectral function,
\ba
\Pi_{ij}(\Omega+i\epsilon)&=&\frac{e^2v_F^2}{2}
\int\limits_{-\infty}^{\infty}
d\omega d\omega^\prime\frac{\tanh\frac{\omega+\mu}{2T}-\tanh
\frac{\omega^\prime+\mu}{2T}}
{\omega-\omega^\prime+\Omega+i\epsilon}\nonumber \\
&\times& \int\frac{d^2k}{(2\pi)^2}\mbox{tr}
\left[\gamma_iA(\omega,\vec{k})\gamma_j
A(\omega^\prime,\vec{k})\right].
\label{electric_cond}
\ea
The vertex corrections were neglected in this expression. Formally, they
are suppressed by a power of $1/N_{f}$. Of course, in the case of graphite
(with $N_{f}=2$), the vertex contributions may nevertheless play an
important role.\cite{D-Lee} This question should be studied in more
detail, but it is outside the scope of the present paper. In absence of a
magnetic field, the conductivity tensor has only the diagonal components,
$\sigma=\sigma_{xx}=\sigma_{yy}$. Both components are equal as a result of
rotational invariance of the model.  The explicit expression of the
conductivity, in this case, reads
\ba
\sigma &=& \frac{e^2N_f}{4\pi^2T}
\int\limits_{-\infty}^\infty
\frac{\Gamma^2 d\omega}{\cosh^2\frac{\omega+\mu}{2T}}
\int\limits_{\Delta^2}^\infty dx
\frac{(x+\omega^2+\Gamma^2)^2-4\omega^2\Delta^2}
{[(x+\omega^2+\Gamma^2)^2-4x\omega^2]^2}\nonumber \\
&=& \frac{e^2N_f}{8\pi^2T}\int\limits_{-\infty}^\infty
\frac{d\omega}{\cosh^2\frac{\omega+\mu}{2T}}
\left[1+\frac{\omega^2-\Delta^2+\Gamma^2}
{2|\omega|\Gamma}\right. \nonumber \\ 
&\times&\left.
\left(\frac{\pi}{2}-\arctan\frac{\Gamma^2+\Delta^2-\omega^2}
{2|\omega|\Gamma}\right)\right],
\label{cond-gamma}
\ea
where $\Gamma$ is the width parameter, and the density of 
carriers is defined by the following relation: 
\ba
n &=& \frac{\Gamma N_f}{2\pi^2 v_F^2}\int\limits_{-\infty}^\infty
\frac{d\omega}{\omega^2+\Gamma^2}\int\limits_{\Delta}^\infty
dEE\nonumber \\
&\times&
\left[\tanh\frac{\omega+\mu+E}{2T}+\tanh\frac{\omega+\mu-E}{2T}\right].
\ea
In the limit $\Gamma\to 0$, these two expressions reduce to
\ba
\sigma&=&\frac{e^2N_f}{16\pi T\Gamma}\int\limits_{-\infty}^\infty
\frac{d\omega}{\cosh^2\frac{\omega+\mu}{2T}}
\frac{\omega^2-\Delta^2}{|\omega|}
\theta(\omega^2-\Delta^2)\nonumber\\
&=&\frac{e^2N_f}{16\pi T\Gamma}\int\limits_{\Delta}^\infty
\frac{d\omega}{\omega}
\left[\frac{\omega^2-\Delta^2}{\cosh^2\frac{\omega+\mu}{2T}}
+(\mu\to -\mu)\right],
\label{cond-no-gamma}
\ea
and
\ba
n &=&\frac{N_f}{2\pi v_F^2}\int\limits_{\Delta}^\infty
dEE\left[\tanh\frac{\mu+E}{2T}+\tanh\frac{\mu-E}{2T}\right]
\nonumber \\
&=&\frac{N_fT^2\sinh\frac{\mu}{T}}{\pi v_F^2}
\int\limits_{\frac{\Delta}{T}}^\infty
\frac{dxx}{\cosh{x}+\cosh\frac{\mu}{T}}
\nonumber \\
&=&\frac{N_fT^2}{\pi v_F^2}\left[\frac{\Delta}{T}
\ln\frac{1+\exp(\frac{\mu-\Delta}
{T})}
{1+\exp(-\frac{\mu+\Delta}{T})}
\right. \nonumber \\
&+&\left.
\mbox{Li}_2\left(-e^{-\frac{\mu+\Delta}{T}}\right)-
\mbox{Li}_2\left(-e^{\frac{\mu-\Delta}{T}}\right)\right],
\ea
where $\mbox{Li}_{2}(z)$ is the dilogarithm function.
As one could see from the above formulas, the conductivity 
grows linearly with temperature when the temperature is large,
\be
\sigma \simeq \frac{e^2N_f}{4\pi} \frac{T}{\Gamma}
\ln 2, \quad \mbox{for} \quad T\to \infty.
\label{sig-largeT}
\ee
Notice, however, that the expression for the conductivity in
Eq.~(\ref{cond-no-gamma}), derived for the $\Gamma\to 0$ case, fails when
temperatures become very small. The correct result for small temperatures
could be derived from Eq.~(\ref{cond-gamma}). It reads
\ba
\sigma &=& \frac{e^2N_f}{2\pi^2}
\left[1+\frac{\mu^2-\Delta^2+\Gamma^2}
{2|\mu|\Gamma}
\right. \nonumber \\&\times&\left.
\left(\frac{\pi}{2}-\arctan\frac{\Gamma^2+\Delta^2-\mu^2}
{2|\mu|\Gamma}\right)+O\left(\frac{T}{\Gamma}\right)\right].
\label{sig-T0}
\ea
The density in that same limit is
\be
n=\frac{N_f}{2\pi v_F^2}(\mu^2-\Delta^2)\mbox{sgn}(\mu)
\theta(\mu^2-\Delta^2).
\label{density-B0-T0}
\ee
The interplay between the density of carriers and the width $\Gamma$
is characterized by the following dimensionless parameter:
\be
\eta = \frac{1}{\Gamma} \sqrt{\frac{2\pi v_F^2 n}{N_f}}.
\ee
In the two opposite limits of a clean or dirty system, this parameter is
either large or small, respectively. Then, the corresponding zero
temperature asymptotes for the conductivity take the form:
\be
\sigma = \frac{e^2N_f\eta}{4\pi}
\sqrt{\frac{2\pi v_F^2 n}{2\pi v_F^2 n +N_{f} \Delta^{2}}},
\ee
for $\eta \gg 1$, and 
\be 
\sigma = \frac{e^2N_f}{2\pi^2}\left[1+\frac{\Gamma}{2\Delta}
\left(\frac{\pi}{2}-\arctan\frac{\Gamma}{2\Delta}\right)\right],
\ee
for $\eta \ll 1$. The last expression was derived under the 
assumption that $n\ll \Delta^{2}/v_{F}^{2}$. Finally, in the 
strict limit of zero density (i.e., $\mu=0$), we derive
\be
\sigma = \frac{e^2N_f}{\pi^2}\frac{\Gamma^{2}}{\Gamma^{2}+\Delta^{2}}.
\label{sig-n0}
\ee
It should be emphasized that the strict case of zero density corresponds
to $\mu=0$ (rather than $\mu=\Delta$ as might be suggested by taking the
limit $T\to 0$ first, and then $n\to 0$). To understand this better, one
should look at the temperature dependence of the chemical potential at a
given fixed value of the density. In particular, when the density of
carriers is very small, the chemical potential as a function of
temperature sharply falls from its value $\mu=\Delta$ at $T=0$ almost down
to zero in a very small region of temperatures. Afterwards, it starts to
grow. When the density gets vanishingly small, the before mentioned region
of temperatures where the chemical potential drops shrinks to zero. Thus,
by making use of continuity argument, it is clear that the value of the
chemical potential is zero in the limit $T\to0$ if the density of carriers
is zero.

It is noticeable that our result in Eq.~(\ref{sig-n0}) is in agreement
with the Wiedemann-Franz law, i.e.,
\be
\left. \frac{\sigma T}{\kappa}\right|_{T\to 0}=\frac{3e^{2}}{\pi^2},
\ee
where we use the value of the thermal conductivity $\kappa$
calculated in Ref.~\onlinecite{GFI}.

\begin{figure}
\epsfxsize=8.0cm
\noindent
\epsffile[90 -5 580 300]{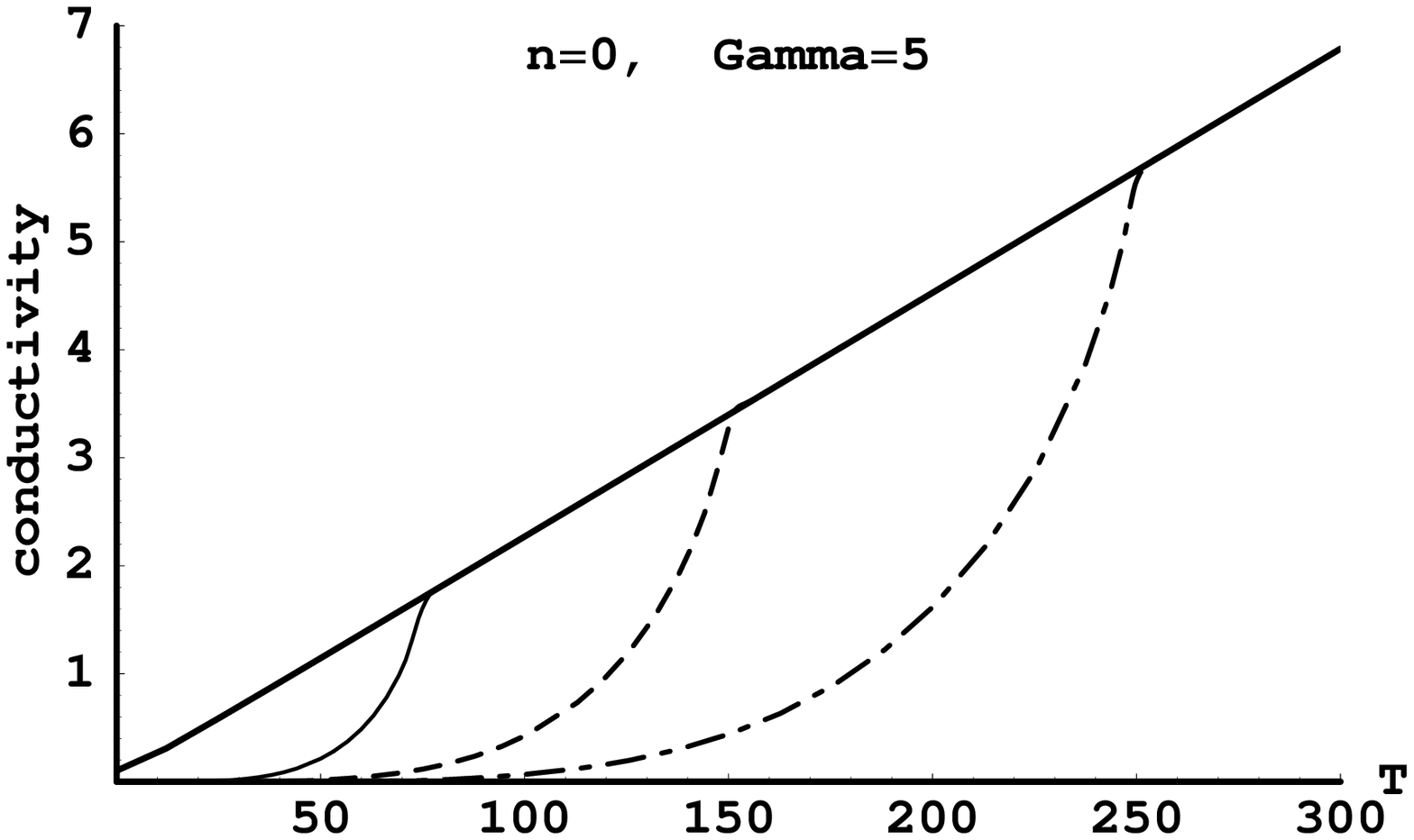}\\
\epsfxsize=8.0cm
\epsffile[90 -5 580 300]{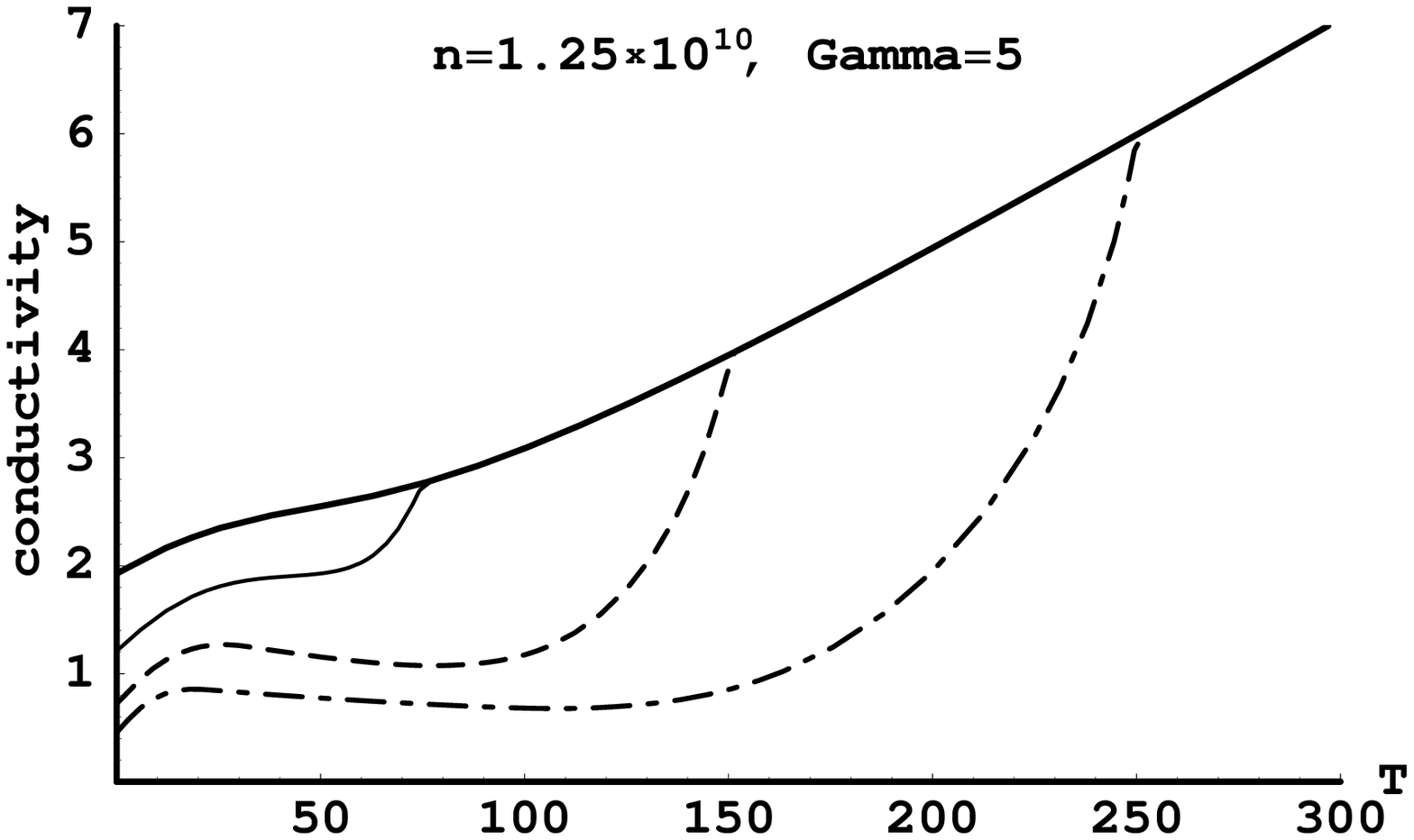}\\
\epsfxsize=8.0cm
\epsffile[90 -5 580 300]{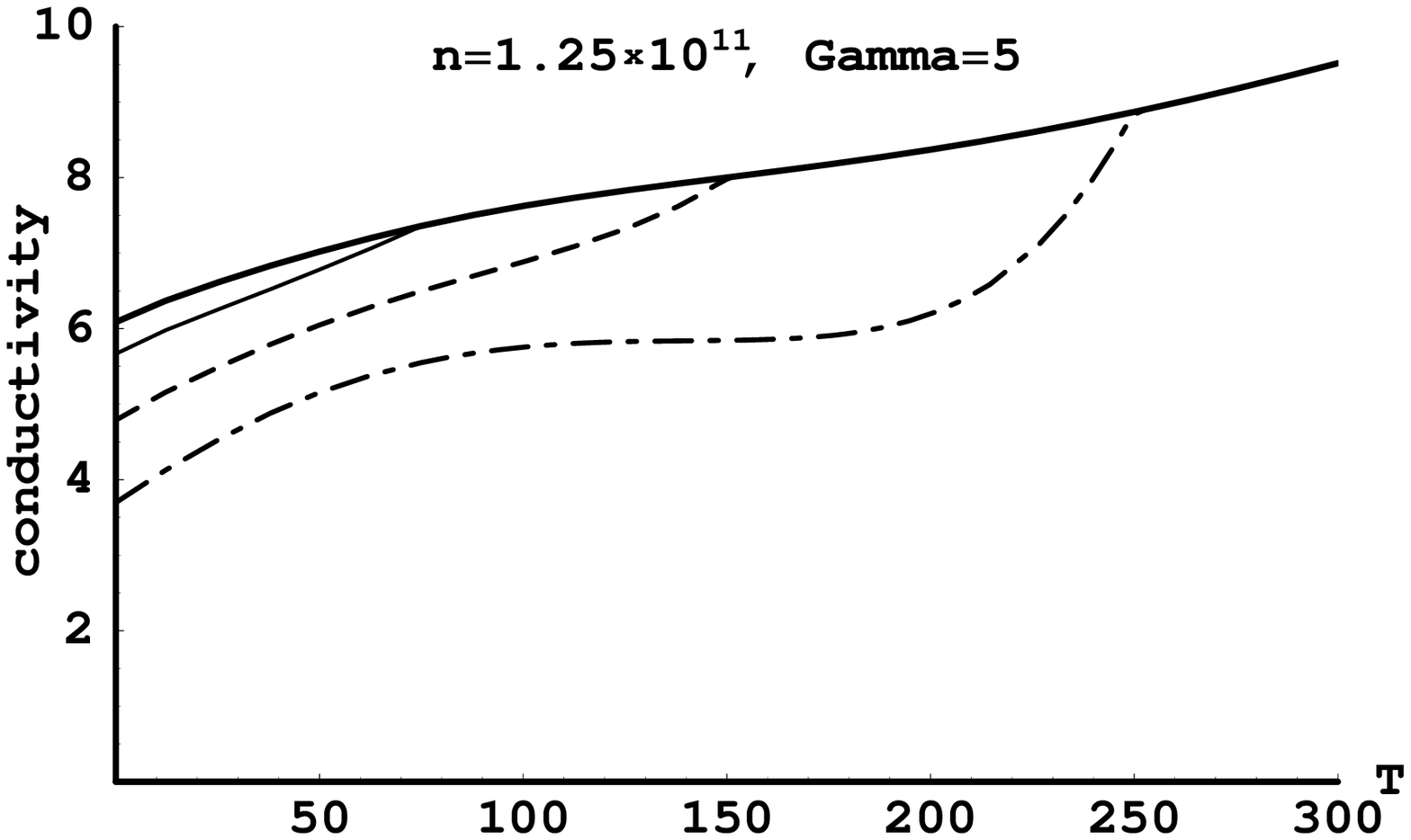}
\caption{The conductivity as a function of temperature for the zero
magnetic field case. The bold solid line corresponds to the case without a
gap. The other lines correspond to nonzero dynamical gaps and different
values of $T_{c}$. Conductivity is measured in units of $e^{2}$, both
temperature and width $\Gamma$ are measured in Kelvin, and density $n$ is
measured in $cm^{-2}$.}
\label{cond-B0}
\end{figure}

The numerical results for the temperature dependence of the conductivity
are shown in Fig.~\ref{cond-B0} in the case of the zero gap (bold solid
line), and nonzero gaps (other lines correspond to different values of
$T_{c}$).  Notice that the model at hand reveals an ``insulator" (i.e.,
increasing with temperature) type behavior of conductivity even in the
case of a finite density of carriers. This type of behavior is the
consequence of using a constant value of the width parameter $\Gamma$ in
our model. Then, the growth of conductivity with increasing temperature is
directly related to the increasing number of thermally excited
quasiparticles.  In realistic systems, of course, the width (which is
related to the inverse scattering time) would normally start to grow with
temperature too. In general, one might choose the width as function of
energy and temperature,
\be
\Gamma(\omega,T) = \Gamma_{0} + \frac{1}{\tau(\omega,T)},
\label{Gamma-T}
\ee
where $\Gamma_{0}$ is the zero temperature width due to impurities, and
the other term is due to the thermal contribution. In this paper, for the
sake of simplicity, we consider the simplest model with a fixed constant
value of the width parameter. The analysis, however, could be easily
generalized for any phenomenologically motivated dependencies like that in
Eq.~(\ref{Gamma-T}).

In order to calculate the conductivities in the case of nonzero dynamical
gaps, we used the gap equation (\ref{gap-Tc}) in which the critical
temperature $T_{c}$ was treated as a free parameter.

\begin{figure}
\epsfxsize=8.0cm
\noindent
\epsffile[90 -5 580 300]{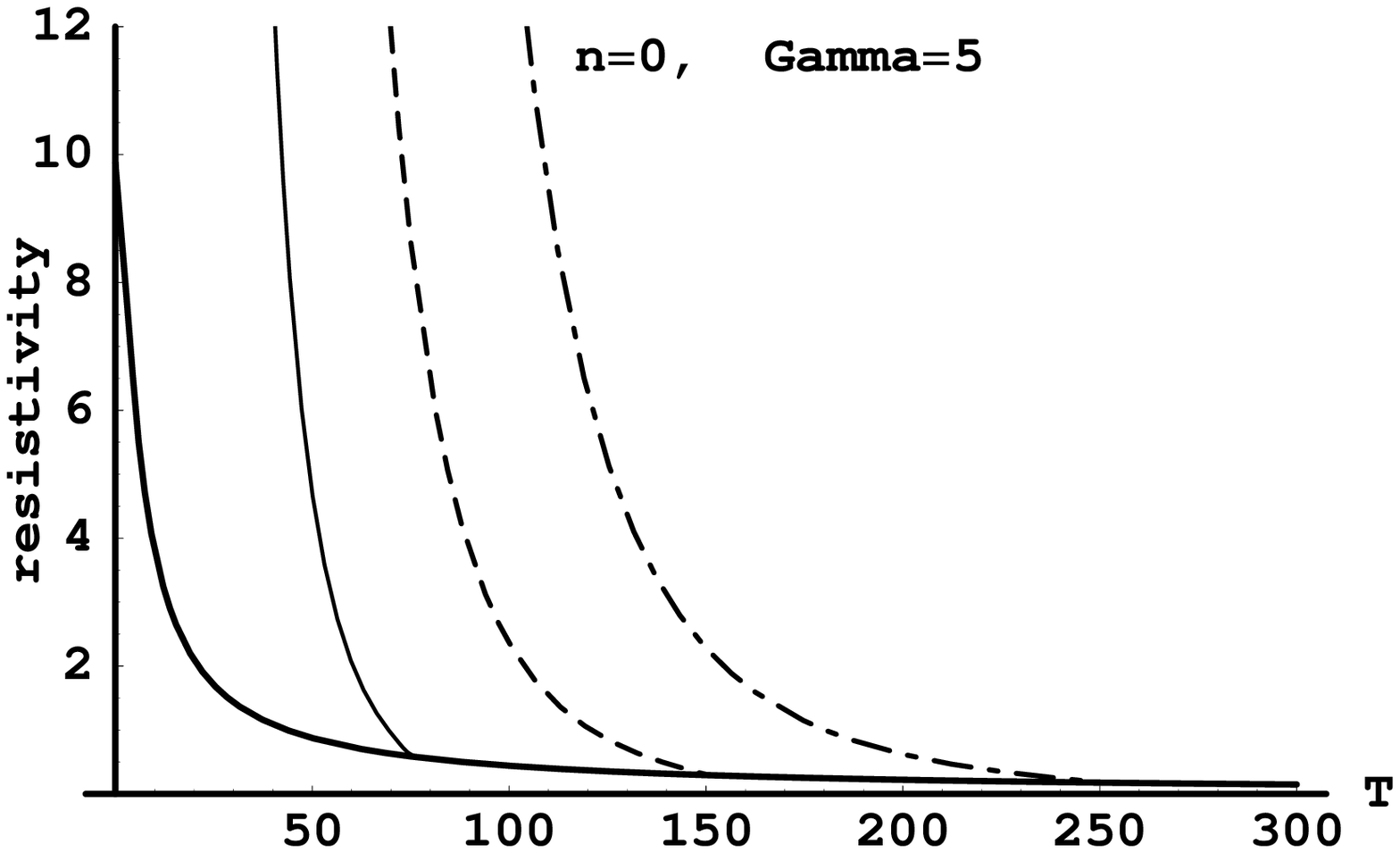}\\
\epsfxsize=8.0cm
\epsffile[90 -5 580 300]{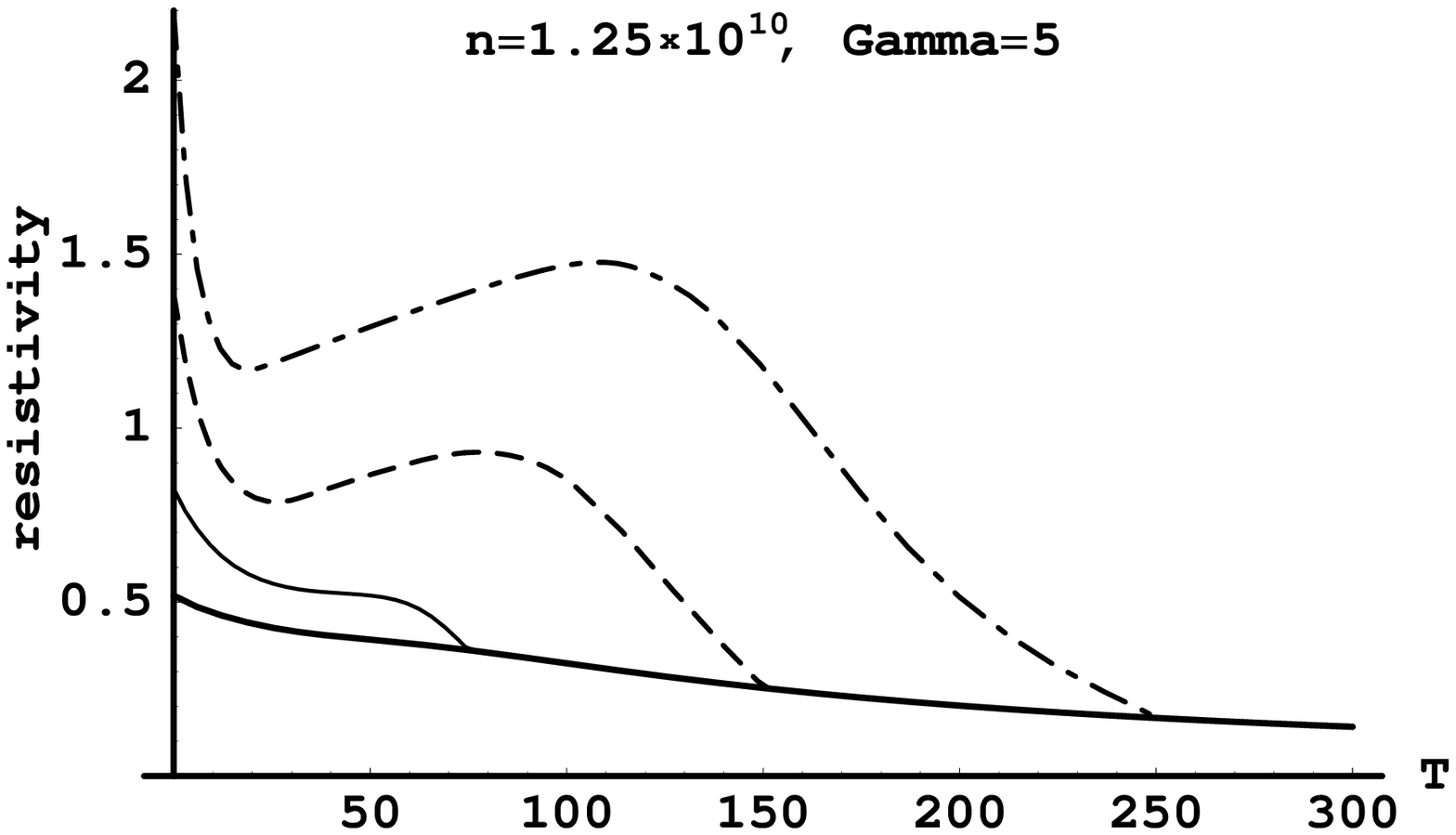}\\
\epsfxsize=8.0cm
\epsffile[90 -5 580 300]{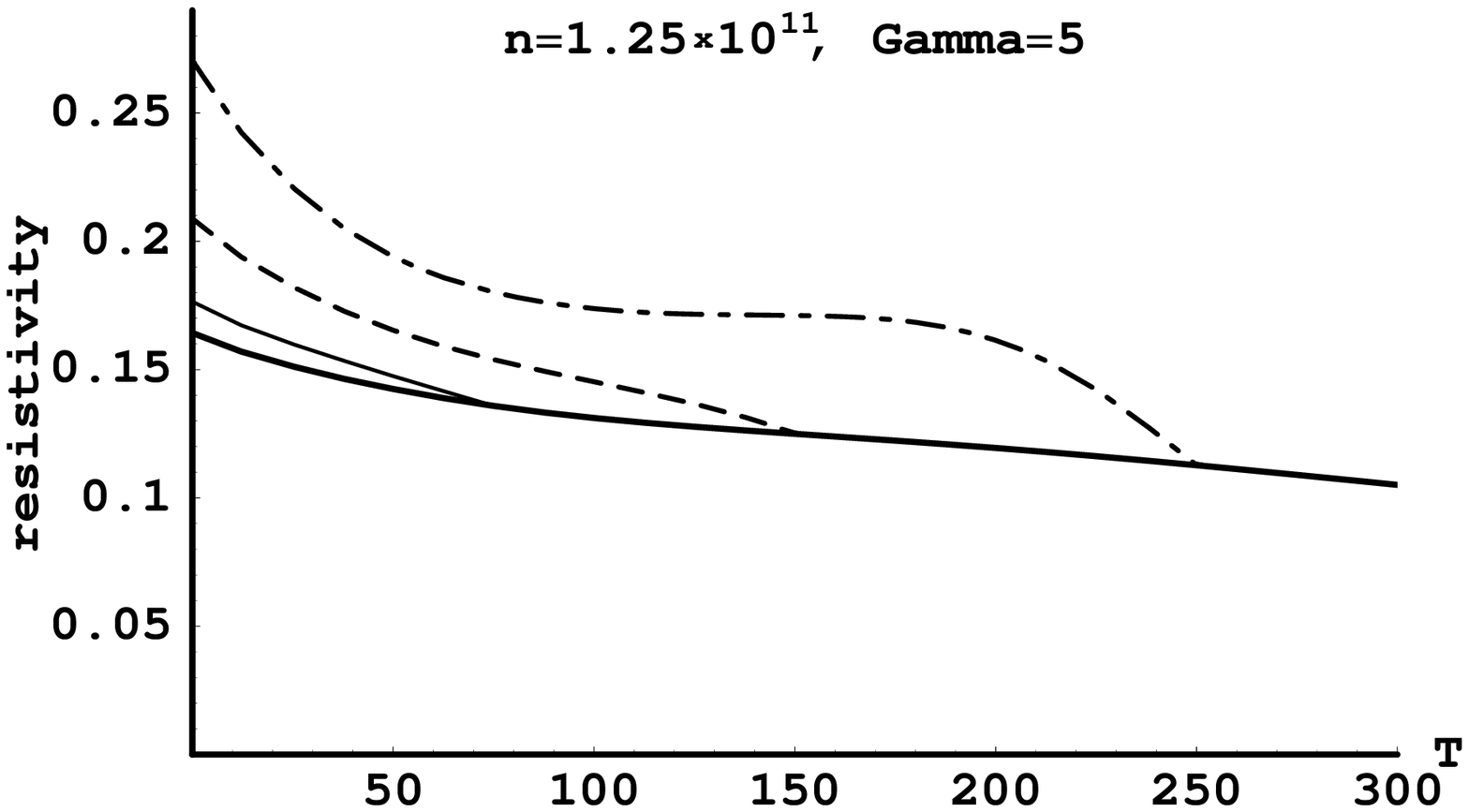}
\caption{The resistivity as a function of temperature for the zero magnetic
field case. The bold solid line corresponds to the case without a gap. The
other lines correspond to nonzero dynamical gaps and different values of
$T_{c}$.  Resistivity is measured in units of $e^{-2}$, both temperature
and width $\Gamma$ are measured in Kelvin, and density $n$ is measured in
$cm^{-2}$.}
\label{resist-B0}
\end{figure}

The results for the temperature dependence of the resistivity are plotted
in Fig.~\ref{resist-B0}.

As one can see in Fig.~\ref{cond-B0} and Fig.~\ref{resist-B0}, there is a
kink at the critical point $T = T_c$ in the conductivity (resistivity). Its
occurrence is directly related to the mean field behavior of the gap
$\Delta$ in the vicinity of the critical point, i.e., $\Delta \sim
\sqrt{T_c - T}$. Indeed, as follows from Eq.~(\ref{cond-gamma}), the
conductivity $\sigma$ depends on $\Delta^2$ and there is a 
linear in $\Delta^2$ term in it as 
$\Delta^2 \to 0$. 
Therefore, its
derivative with respect to temperature has a finite discontinuity at the
critical point $T = T_c$.
The mean field behavior may change if higher order, $1/N_{f}$, corrections
(fluctuations) are taken into account. The fluctuations could either
change the phase transition to a first order one, with a discontinuity in
$\Delta$ at the phase transition point, or to a 
non-mean-field second order phase
transition, with the scaling law $\Delta \sim (T_c - T)^{\nu}$ where $\nu
> 1/2$. While in the former case a discontinuity will appear in the
conductivity (resistivity), in the latter case the conductivity
(resistivity) will be a smooth function of temperature, and a singularity
will move to its higher derivatives.
Another noticeable point is that in the case with no magnetic field, the
flavor phase transition does {\it not} look as a semimetal-insulator one.
Indeed, as one can see in in Fig.~\ref{cond-B0} and Fig.~\ref{resist-B0},
an ``insulator" type behavior does not change at the critical point. As we
will see in the next subsections, the occurrence of a magnetic field will
drastically change this feature of the phase transition.

Before concluding this section, let us also mention that the conductivity
(as well as the resistivity) become more sensitive to the appearance of a
dynamical gap when the density of carriers decreases. To support this
statement, we plotted the conductivity and resistivity for two different
finite values of carrier densities which differ by a factor $10$, see
lower parts of Figs.~\ref{cond-B0} and \ref{resist-B0}.

As we see from Fig.~\ref{resist-B0}, the temperature dependence of the
resistivity develops a minimum when the value of the gap is sufficiently
large. Comparing this temperature dependence with the experimental data,
we might even suggest that the studied graphite samples are better
described by the model with a nonzero dynamical gap even in absence of a
magnetic field. The effect of an external field is studied in the
following subsections.

\subsection{Conductivity tensor. Nonzero magnetic field.}
\label{sec:cond-mag}

Let us now turn to the analysis of the conductivity in the case with an
external magnetic field.  The spectral function $A(\omega ,{\vec{k }})$ of
the translation invariant part of the quasiparticle propagator in a
magnetic field [see Eqs.~(\ref{G-Fourier}) and (\ref{G-LLs})] is given by
\ba
A(\omega,{\vec{k}}) &=& \frac{\Gamma }{\pi}
\exp\left(-\frac{c|\vec{k}|^{2}}{|eB|}\right)
\sum\limits_{n=0}^{\infty} \frac{(-1)^{n}}{M_{n}}
\nonumber \\
&\times& \left[
\frac{(\gamma^{0}M_{n}+\Delta)f_{1}({\vec{k}})+f_{2}({\vec{k}})}
{(\omega-M_{n})^{2}+\Gamma ^{2}}
\right.\nonumber \\
&+&\left.
\frac{(\gamma ^{0}M_{n}-\Delta )f_{1}({\vec{k}})-f_{2}({\vec{k}})}
{(\omega +M_{n})^{2}+\Gamma ^{2}}\right] ,
\label{spect-fun-magfield}
\ea
where $M_n=\sqrt{\Delta^{2}+2nv_{F}^{2}|eB|/c}$ and 
the functions $f_{1}({\vec{k}})$ and $f_{2}({\vec{k}})$ were 
defined earlier in Eqs.~(\ref{expr:f1}) and (\ref{expr:f2}). 
In an external magnetic field, the conductivity is a tensor 
quantity. The diagonal and off-diagonal components of conductivity 
read
{\small
\ba
&&\sigma_{xx} = \frac{e^{2}N_{f}|eB|\Gamma^{2}}{2\pi^{2}T}
\sum_{n=0}^{\infty}\int_{-\infty}^{\infty} 
\frac{d\omega}{\cosh^{2}\frac{\omega+\mu}{2T}} \times \nonumber \\
&& \frac{(\omega^{2}+M_{n}^{2}+\Gamma^{2})
(\omega^{2}+M_{n+1}^{2}+\Gamma^{2})-4\omega^{2}\Delta^{2}}
{\left[(\omega^{2}-M_{n}^{2}-\Gamma^{2})^{2}
+4\omega^{2}\Gamma^{2}\right]
\left[(\omega^{2}-M_{n+1}^{2}-\Gamma^{2})^{2}
+4\omega^{2}\Gamma^{2}\right]}, \nonumber \\
\label{cond-B}
\ea
}
and 
\be
\sigma_{xy} = \frac{e^{2}N_{f}}{2\pi} \nu_{B},
\label{sigma-xy}
\ee
respectively. Here the parameter $\Gamma$ gives the energy width of Landau
levels, and the filling factor $\nu_{B}$ [related to the density of
carriers by the relation:  $\nu_{B}=2\pi c n/(N_{f}|eB|)$] is defined as
follows:
\ba
\nu_{B} &=& \int_{-\infty}^{\infty} \frac{d\omega}{2\pi}
\tanh\frac{\mu+\omega}{2T}\left[\frac{\Gamma}
{(\omega-\Delta)^{2}+\Gamma^{2}}+(\omega\to -\omega)
\right.\nonumber \\
&+& 2 \sum_{n=1}^{\infty}\left.\left(
\frac{\Gamma}{(\omega-M_{n})^{2}+\Gamma^{2}}+(\omega\to -\omega)
\right)\right].
\ea
The sum over Landau levels in Eq.~(\ref{cond-B}) could be performed 
explicitly, and the result is given in terms of the digamma
function, $\psi(z)$, as follows:
\ba
&&\sigma_{xx} = \frac{e^{2}N_{f}\Gamma}{4\pi^{2}T}
\int_{-\infty}^{\infty} \frac{d\omega}{\cosh^{2}\frac{\omega+\mu}{2T}}
\frac{\Gamma}{(\frac{v_{F}^{2}eB}{c})^{2}+(2\omega\Gamma)^{2}} 
\Bigg[2\omega^{2}
\nonumber \\
&+& \frac{(\omega^{2}+\Delta^{2}+\Gamma^{2})
\left(\frac{v_{F}^{2}eB}{c}\right)^{2}
-2\omega^{2}(\omega^{2}-\Delta^{2}+\Gamma^{2})\frac{v_{F}^{2}eB}{c}}
{(\omega^{2}-\Delta^{2}-\Gamma^{2})^{2}+4\omega^{2}\Gamma^{2}}
\nonumber \\
&-& \left.\frac{\omega(\omega^{2}-\Delta^{2}+\Gamma^{2})}{\Gamma}
\mbox{Im~}\psi\left(\frac{\Delta^{2}+\Gamma^{2}-\omega^{2}-2i\omega\Gamma}
{2v_{F}^{2}|eB|/c}
\right)\right]. \nonumber \\
\label{cond-B-sum}
\ea
The high temperature asymptote that follows from the representation in
Eq.~(\ref{cond-B-sum}) is the same as in the case of zero magnetic field,
given in Eq.~(\ref{sig-largeT}). The limit $T\to 0$ is different from that
in Eq.~(\ref{sig-T0}). It is given by the following expression:
\ba
&&\sigma_{xx} = \frac{e^{2}N_{f}\Gamma}{\pi^{2}}
\frac{\Gamma}{(\frac{v_{F}^{2}eB}{c})^{2}+(2\mu\Gamma)^{2}}
\Bigg[2\mu^{2}
\nonumber \\
&+& \frac{(\mu^{2}+\Delta^{2}+\Gamma^{2})
\left(\frac{v_{F}^{2}eB}{c}\right)^{2}
-2\mu^{2}(\mu^{2}-\Delta^{2}+\Gamma^{2})\frac{v_{F}^{2}eB}{c}}
{(\mu^{2}-\Delta^{2}-\Gamma^{2})^{2}+4\mu^{2}\Gamma^{2}}
\nonumber \\
&+& \left.\frac{\mu(\mu^{2}-\Delta^{2}+\Gamma^{2})}{\Gamma}\mbox{Im~}
\psi\left(\frac{\Delta^{2}+\Gamma^{2}-\mu^{2} + 2i\mu\Gamma}
{2v_{F}^{2}|eB|/c}
\right)\right].\nonumber \\
\ea
It is interesting to note, however, that for zero value of the gap
and zero density of carriers (i.e., $\Delta=0$ and $\mu=0$), this last
expression becomes identical with the expression for the conductivity
in absence of a magnetic field given in Eq.~(\ref{sig-n0}).

In the limit of narrow width, $\Gamma\to 0$, the above expressions
reduce down to
\ba
\sigma_{xx} &=& \frac{e^{2}N_{f}\Gamma}{2\pi T} \left[
\frac{1+\cosh\frac{\Delta}{T}\cosh\frac{\mu}{T}}
{(\cosh\frac{\Delta}{T}+\cosh\frac{\mu}{T})^2}\right.
\nonumber \\
&+& 4\sum_{n=1}^{\infty}\left.
\frac{n(1+\cosh\frac{M_{n}}{T}\cosh\frac{\mu}{T})}
{(\cosh\frac{M_{n}}{T}+\cosh\frac{\mu}{T})^2}\right] ,
\ea
for diagonal component of the conductivity, and 
\ba
\nu_{B} &=& \frac{1}{2}\left(\tanh
\frac{\mu+\Delta}{2T}+\tanh\frac{\mu-\Delta}{2T}\right) \nonumber \\
&+&\sum\limits_{n=1}^\infty
\left(\tanh\frac{\mu+M_n}{2T}+\tanh\frac{\mu-M_n}{2T}\right),
\ea
for the filling factor. 

\begin{figure}
\epsfxsize=8.0cm
\epsffile[90 -5 580 300]{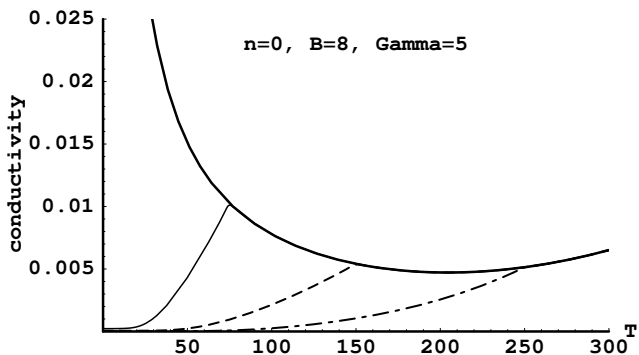}
\caption{The diagonal component of conductivity as a function of
temperature for zero value of carrier density and a nonzero magnetic
field, $B=8$ Tesla. The bold solid line corresponds to the case without a
gap. Other lines correspond to nonzero dynamical gaps and different values
of $T_{c}$. Conductivity is measured in units of $e^{2}$, both temperature
and width $\Gamma$ are measured in Kelvin.}
\label{cond-nu0}
\end{figure}

In order to understand the effect of a dynamical gap on the behavior of
conductivity as a function of temperature, it is helpful to start from the
case of a vanishing density of carriers (i.e., $\nu_{B}=0$).  When
$\nu_{B}=0$, the Hall conductivity is absent, and the resistivity is
determined by $\sigma_{xx}$ component alone. The plot of the conductivity
as a function of temperature is given in Fig.~\ref{cond-nu0}.
The bold solid line corresponds to the case without a dynamical gap, while
the other lines correspond to different choices of the gap magnitude. In
the numerical analysis, we used the gap equation in Eq.~(\ref{gap-Tc}),
keeping the value of $T_{c}$ as a free parameter.

When there are no free carriers, the low temperature dependence of the
diagonal component of conductivity is very sensitive to the presence of a
gap. In absence of a gap, the conductivity becomes infinitely large when
$T\to 0$. At the same time, it is zero in the same limit when there is
even an arbitrarily small gap $\Delta$.

An important fact is that, unlike the case without magnetic field, the
conductivity exhibits a (semi)metallic type behavior for zero gap and not
too high temperatures ($T \lesssim 200$ K in Fig.~\ref{cond-nu0}). As a
result, for not too large values of the critical temperature $T_c$ ($T_{c}
\lesssim 200$ K in Fig.~\ref{cond-nu0}), the flavor phase transition looks
as a conventional semimetal-insulator one, when the insulator type
behavior below $T_c$ (nonzero gap) is replaced by the metallic type in a
range of temperatures just above $T_c$ (zero gap) (see
Fig.~\ref{cond-nu0}).

\begin{figure}
\epsfxsize=8.0cm
\noindent
\epsffile[90 -5 580 300]{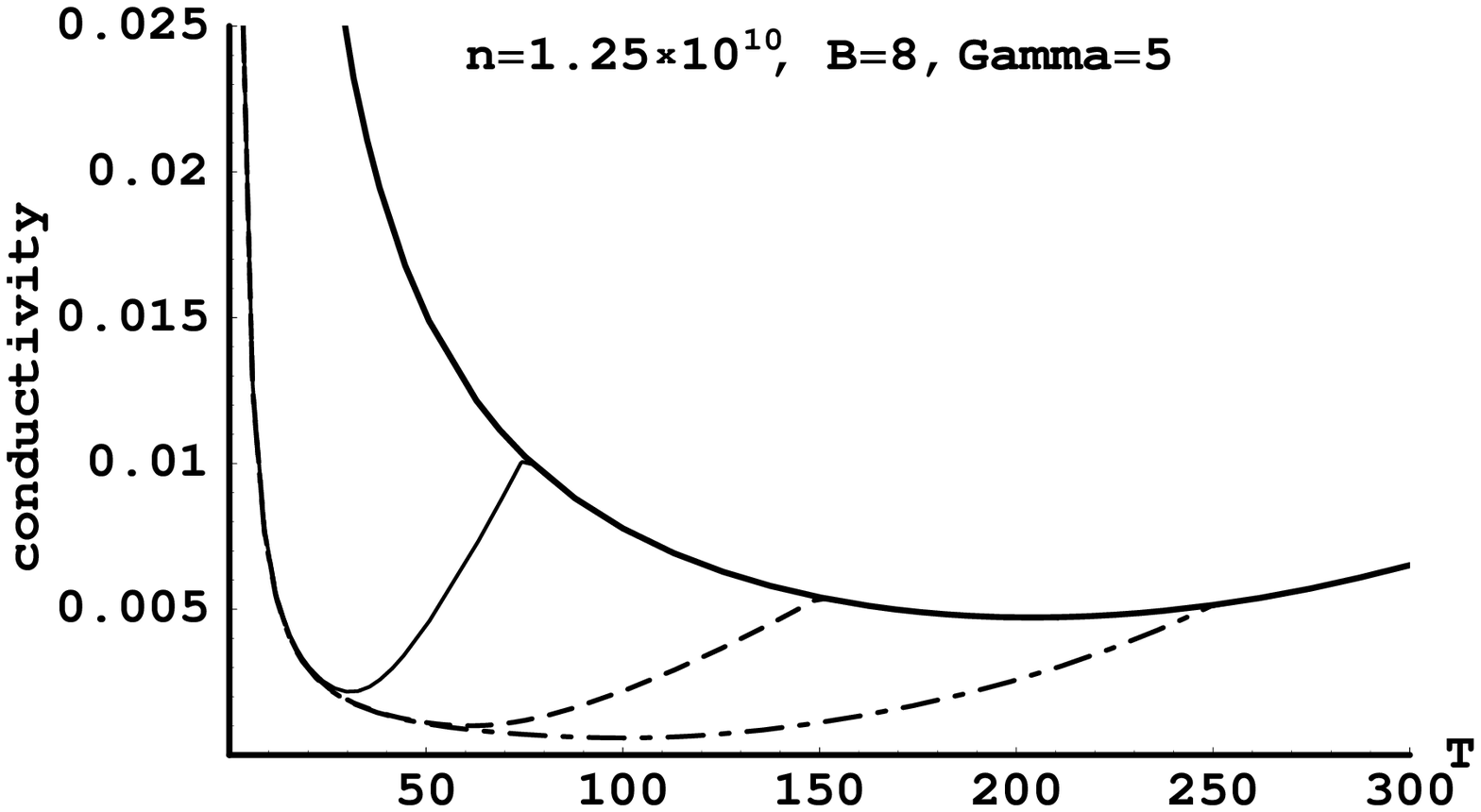}\\
\epsfxsize=8.0cm
\epsffile[90 -5 580 300]{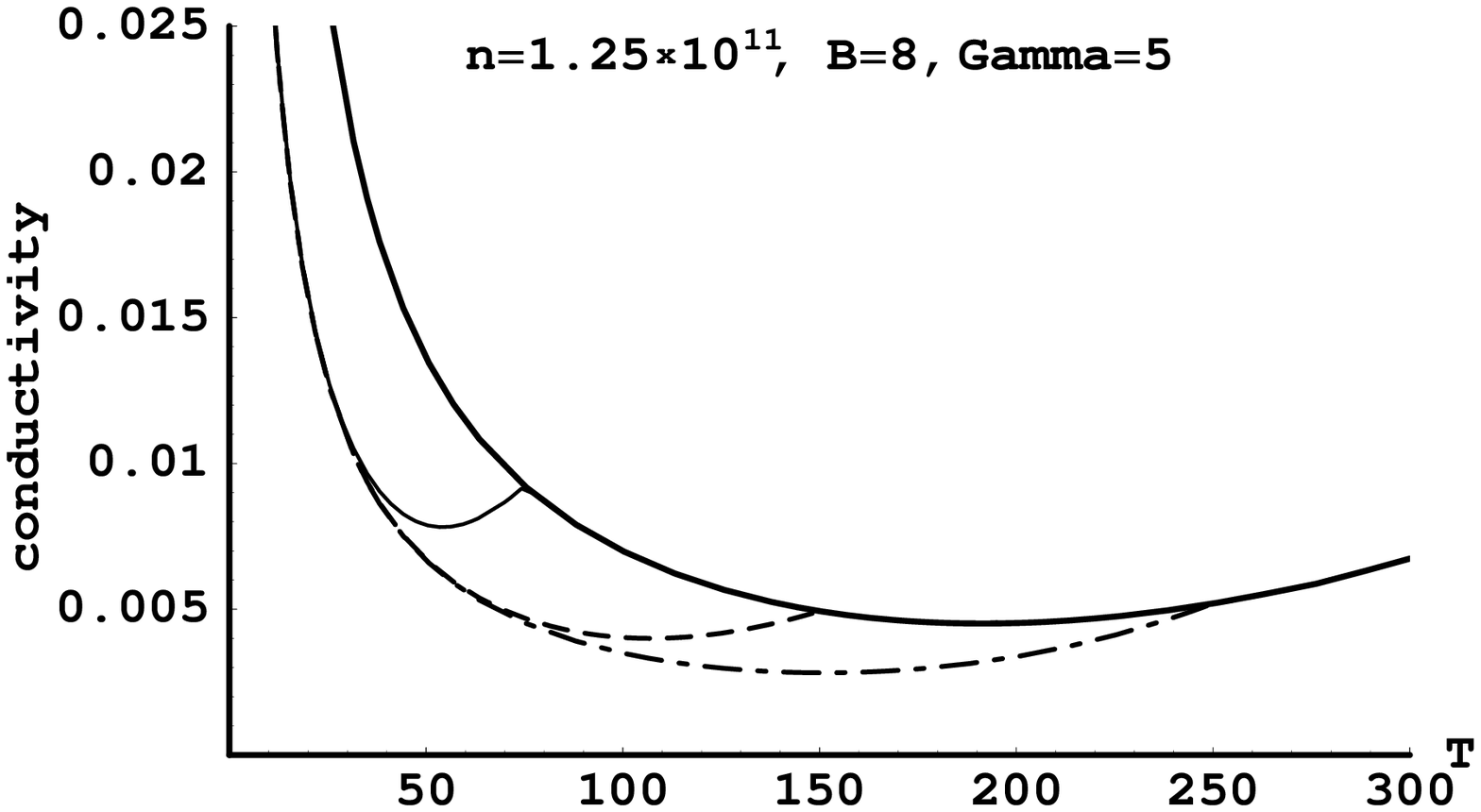}
\caption{The diagonal component of conductivity as a function of
temperature for two different densities and a nonzero magnetic field,
$B=8$ Tesla. The bold solid line corresponds to the case without a gap.
Other lines correspond to nonzero dynamical gaps and different values of
$T_{c}$. Conductivity is measured in units of $e^{2}$, both temperature
and width $\Gamma$ are measured in Kelvin, and density $n$ is measured in
$cm^{-2}$.}
\label{cond-nu}
\end{figure}

A typical conductivity for a nonzero value of the filling factor $\nu_{B}$
(i.e., nonzero charge density) is shown in Fig.~\ref{cond-nu}. In this
case, the behaviors of the conductivity for a nonzero gap and zero gap are
more similar than in the case of $\nu_{B} = 0$ (compare with
Fig.~\ref{cond-nu0}). The presence of a gap, however, can substantially
reduce the value of conductivity in the whole range of temperatures below
$T_{c}$. It is important that, like in the case with $\nu_{B} = 0$, the
flavor phase transition looks as a semimetal-insulator one for not too
large values of the critical temperature $T_c$ (see Fig.~\ref{cond-nu}).
It is the most important conclusion of this subsection.

The same arguments as in the end of Sec.~\ref{nomag} show that the
occurrence of the kink in the conductivity at $T = T_c$ reflects the 
mean-field behavior of the gap in the vicinity of the critical point,
$\Delta
\sim \sqrt{T_c - T}$. The $1/N_f$ fluctuations may change the character of
the phase transition, leading either to a discontinuity in the
conductivity $\sigma(T)$ at $T = T_c$ (a first order phase transition) or
to a smooth function $\sigma(T)$ (a non-mean-field second order phase
transition).

\subsection{Resistivity tensor}

In this subsection, we study the temperature dependence 
of the resistivity.

In terms of conductivities, the diagonal component of the 
resistivity reads
\be
\rho_{xx} = \frac{\sigma_{xx}}{\sigma_{xx}^{2}+\sigma_{xy}^{2}}.
\label{resist-xx}
\ee
In order to understand the general behavior of the resistivity,
below we perform a set of numerical calculations. 

Before presenting the results, it is instructive to notice that there
exist two opposite regimes of dynamics controlled by the value of the
charge density. In particular, at small density, when the Hall
conductivity $\sigma_{xy}$ is negligible compared to $\sigma_{xx}$, the
resistivity in Eq.~(\ref{resist-xx}) behaves as $1/\sigma_{xx}$. On the
other hand, at sufficiently large density, when the Hall conductivity
dominates over the diagonal component, the resistivity $\rho_{xx} \approx
\sigma_{xx}/\sigma_{xy}^{2}$. By recalling that the Hall conductivity [see
Eq.~(\ref{sigma-xy})] is independent of temperature, the general features
of the temperature dependence of $\rho_{xx}$ will be the same as of
$1/\sigma_{xx}$ and $\sigma_{xx}$ in the mentioned two regimes,
respectively.

\begin{figure}
\epsfxsize=8.0cm
\epsffile[90 -5 580 300]{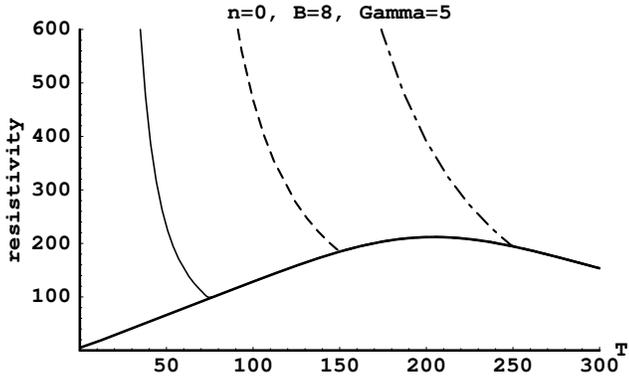}
\caption{The resistivity as a function of temperature for zero value of
carrier density and a nonzero magnetic field, $B=8$ Tesla. The bold solid
line corresponds to the case without a gap. Other lines correspond to
nonzero dynamical gaps and different values of $T_{c}$. Resistivity is
measured in units of $e^{-2}$, both temperature and width $\Gamma$ are
measured in Kelvin.}
\label{resist-nu0}
\end{figure}

Now, let us present the numerical results. We begin by considering the
case of zero density. In this case the Hall conductivity equals zero and
the resistivity $\rho_{xx}$ equals $1/\sigma_{xx}$.  The temperature
dependence of the resistivity is shown in Fig.~\ref{resist-nu0} (compare
with Fig.~\ref{cond-nu0}).
The bold solid line corresponds to a model with the vanishing dynamical
gap in the quasiparticle spectrum. The other curves correspond to three
different choices of the dynamical gap. The temperature dependence of the
dynamical gap is given by the gap equation (\ref{gap-Tc}) where the value
of $T_{c}$ was treated as a phenomenological parameter.

As one can see in Fig.~\ref{resist-nu0}, for zero dynamical gap (the bold
solid line) the resistivity has a metallic type behavior for not too high
temperatures ($T\lesssim 0.2 v_{F}\sqrt{|eB|/c}$) and an insulator type
behavior at high temperatures ($T\gtrsim 0.2 v_{F} \sqrt{|eB|/c}$). This
type of temperature dependence is driven by the magnetic field alone and
is not related to the generation of a dynamical gap. Such a crossover from
the metallic type behavior (low temperatures) to the insulator one (high
temperatures) can be easily distinguished from the flavor phase transition
taking place at not too high $T_c$. Indeed, when $T_{c} \lesssim 0.2
v_{F}\sqrt{|eB|/c}$, one can see from Fig.~\ref{resist-nu0} that it
corresponds to the opposite, conventional, transition, when the insulator
type behavior below $T_c$ (nonzero gap) is replaced by the metallic type
in a range of temperatures just above $T_c$ (zero gap).

\begin{figure}
\epsfxsize=8.0cm
\noindent
\epsffile[90 -5 580 300]{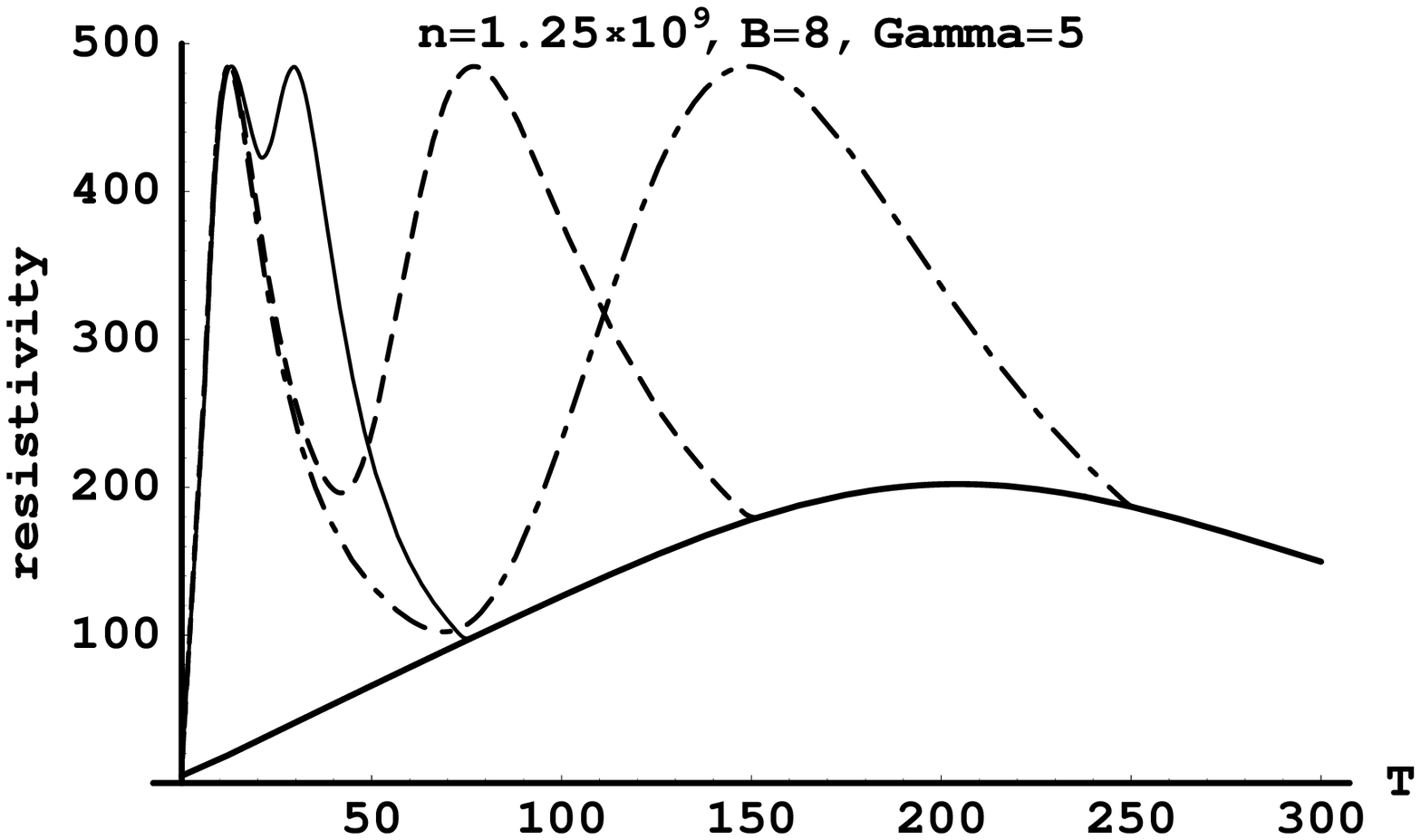}\\
\epsfxsize=8.0cm
\epsffile[90 -5 580 300]{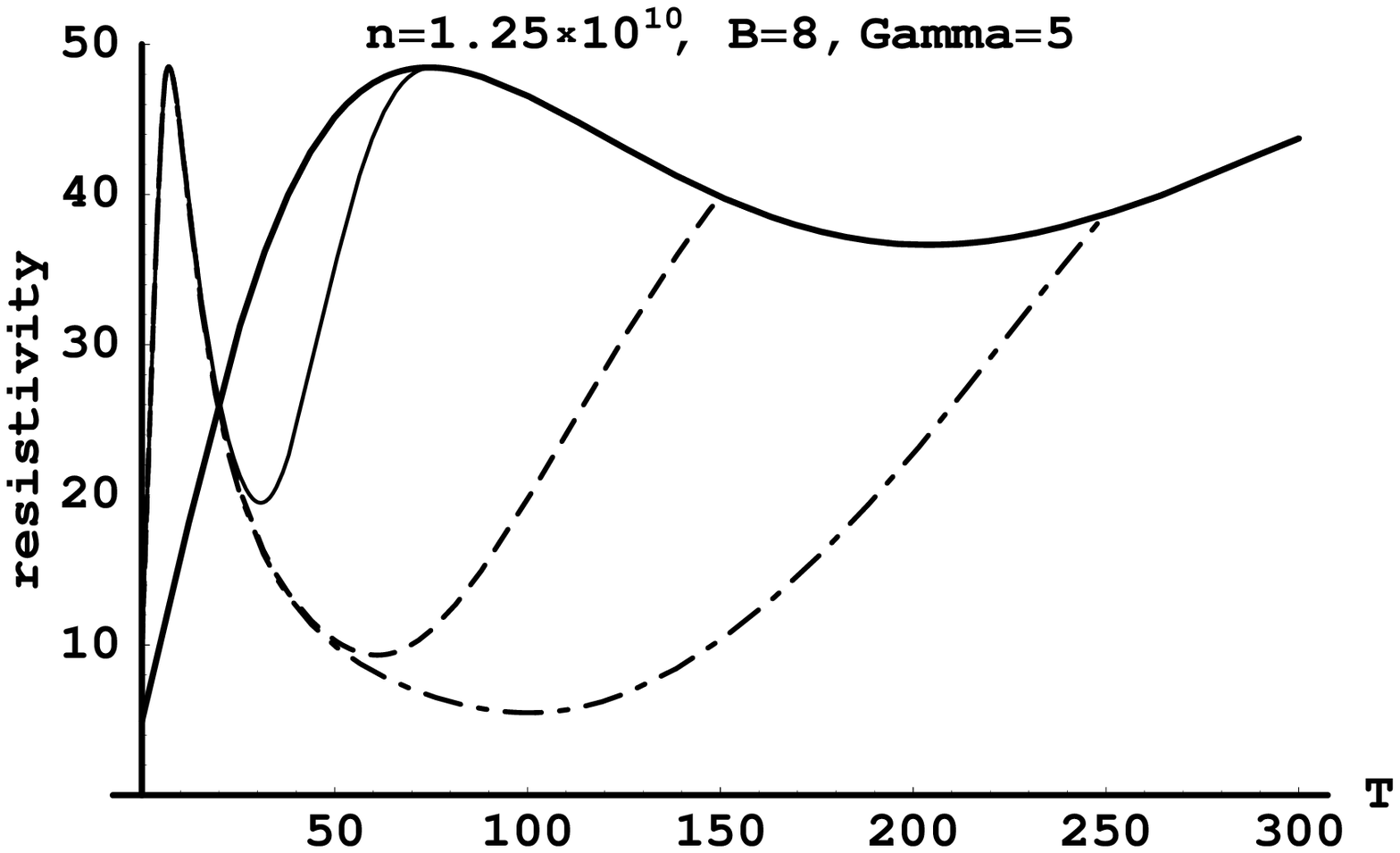} \\
\epsfxsize=8.0cm
\epsffile[90 -5 580 300]{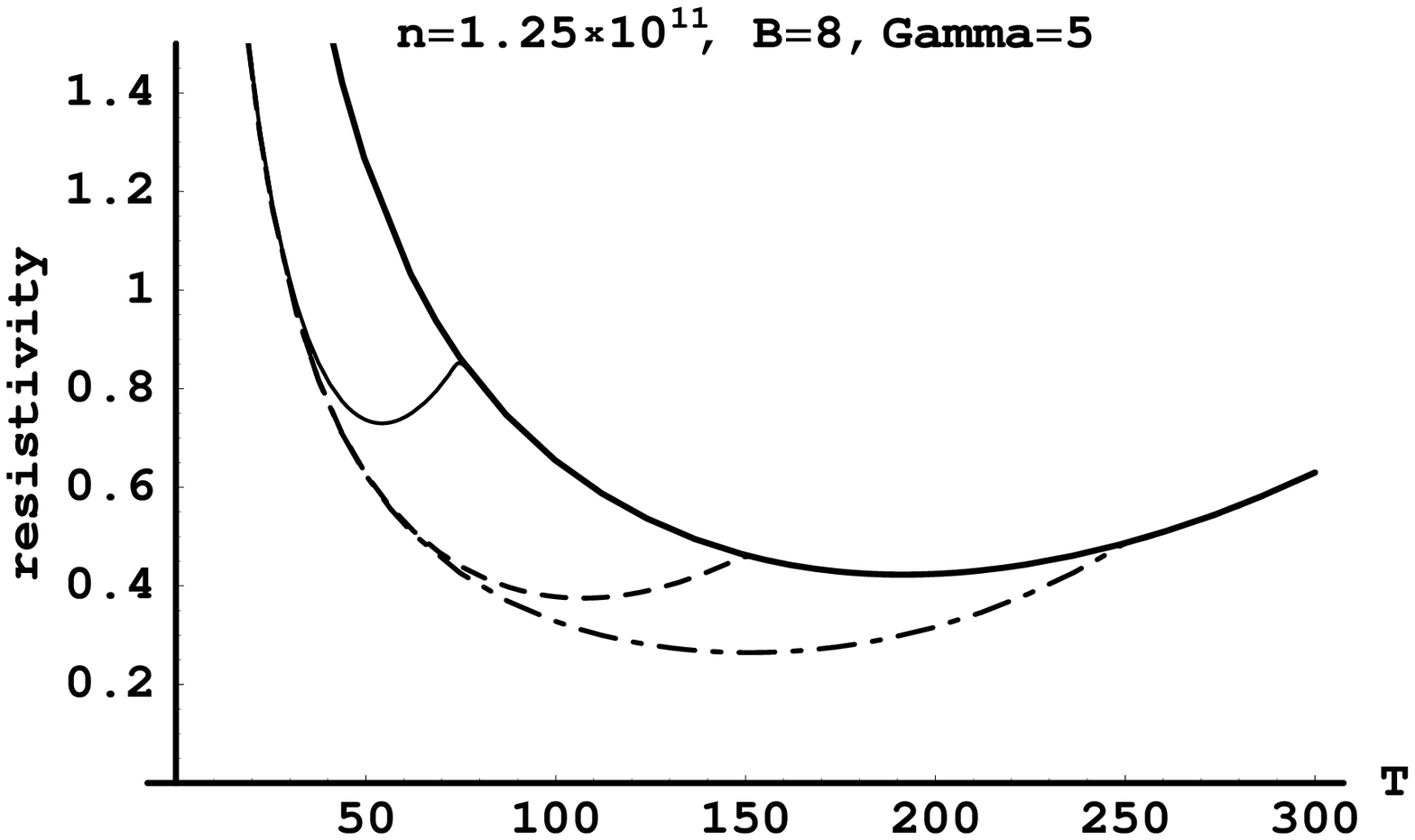}
\caption{The diagonal component of resistivity as a function of temperature
for four different densities and a nonzero magnetic field, $B=8$ Tesla.
The bold solid line corresponds to the case without a gap. Other lines
correspond to nonzero dynamical gaps and different values of $T_{c}$.
Resistivity is measured in units of $e^{-2}$, both temperature and width
$\Gamma$ are measured in Kelvin, and density $n$ is measured in
$cm^{-2}$.}
\label{resist-nu}
\end{figure}

Let us now proceed to the case of a nonzero charge density. As we
mentioned at the beginning of this subsection, there are two different
regimes that appear in the limits of small and large densities,
respectively. Our results in Fig.~\ref{resist-nu} illustrate these
regimes. 
At small density (see the upper panel in Fig.~\ref{resist-nu}),
the resistivity behaves as $1/\sigma_{xx}$ almost at all temperatures
(compare with Fig.~\ref{cond-nu}), except for a finite region where
$\sigma_{xx}$ becomes very small due to the generation of a dynamical gap.
In this region, even a small value of the Hall conductivity could dominate
over $\sigma_{xx}$. This is seen as the appearance of a local minimum
(between two maxima) in the temperature dependence of the resistivity.  Of
course, in the absence of a dynamical gap (see the solid line in the upper
panel in Fig.~\ref{resist-nu}), the resistivity remains essentially the
same as that at $n=0$ (compare with Fig.~\ref{resist-nu0}).

This picture changes dramatically with increasing the density. As one can
see in the two lower panels in Fig.~\ref{resist-nu}, with increasing the
density the resistivity gradually approaches the regime where $\rho_{xx}
\sim \sigma_{xx}$.

For critical temperatures $T_{c} \lesssim 0.2 v_{F}\sqrt{|eB|/c}$ (i.e.,
for dependencies represented by thin solid and dashed lines), these two
regimes correspond to two essentially different metal-insulator phase
transitions. In the case of a small density (see the upper panel in
Fig.~\ref{resist-nu}), it is a conventional phase transition with
insulator type and metallic type behaviors at temperatures below and just
above $T_c$, respectively. On the other hand, for a large density (see the
lower panel in Fig.~\ref{resist-nu}), the ``inverse" metal-insulator phase
transition (with metal type dependence just below $T_c$ and insulator type
just above $T_c$) is realized.

Therefore we conclude that, in the presence of a magnetic field, a
dynamical gap in the quasiparticle spectrum can indeed lead to a change of
the insulator type dependence of $\rho_{xx}(T)$ to the metallic one.
However, the nonzero Hall conductivity at finite density $n$ complicates
the picture and can lead to different types of metal-insulator phase
transitions for small and large values of the charge density.

When the metal-insulator phase transition is truly a mean-field one,
its clear signature is a kink in the resistivity $\rho_{xx}(T)$ at the
critical point $T = T_c$. As has been already pointed out in the previous
subsections, the $1/N_{f}$ fluctuations can change this feature, leading
either to a discontinuity in the resistivity at $T = T_c$ (a first order
phase transition) or to a smooth function $\rho_{xx}(T)$ (a 
non-mean-field continuous phase transition).

\section{Metal-insulator phase transition in highly oriented pyrolytic
graphite}
\label{HOPG}

The main motivation of this study was the experimental data reported in
Refs.~\onlinecite{Exp1,SSCom115,Exp2}. It was observed that samples of
highly oriented pyrolytic graphite in an external magnetic field show a
qualitative change of their resistivity as a function of temperature, that
was interpreted as a metal-insulator phase transition. The effect is
clearly seen only for a magnetic field perpendicular to the basal plane,
suggesting that the orbital motion of quasiparticles is responsible for
the change of the conductivity dependence.

In this section we will attempt to explain qualitatively the main features
of the above mentioned experimental data in the light of the magnetic
catalysis idea. We should note that the first step in this direction was
made in Ref.~\onlinecite{Khvesh}. Here we go into further details
utilizing the rather complete description of the magnetic catalysis in
planar systems and its effect on the temperature dependence of their
conductivity and resistivity given in the previous sections.

First of all, the analysis made in Sec.~\ref{condres} shows that, in the
presence of a magnetic field, the flavor phase transition in planar
systems can indeed manifest itself as a metal-insulator phase transition
in the behavior of their resistivity $\rho(T)$ as a function of
temperature. A noticeable fact is the existence of clearly distinguishable
signatures of different types of the phase transition: the presence of a
discontinuity and a kink in the resistivity $\rho(T)$ at the critical
point $T = T_c$ in the cases of first order and mean-field phase
transitions, respectively, and a smooth behavior of $\rho(T)$ at $T = T_c$
for a non-mean-field 
continuous phase transition. To the best of our
knowledge, so far there have been no experiments reporting observations of
a singular behavior of $\rho(T)$ at the critical point. At this stage,
however, it would be premature to conclude that the observed phase
transition is a continuous non-mean-field one. This point deserves further
experimental study.

A very interesting experimental observation made in
Refs.~\onlinecite{Exp1,SSCom115,Exp2} (and, to the best of our knowledge,
has not been explained)  is the existence of a finite ``offset" magnetic
field $B_{c}$. The value $B_c$ determines the threshold $B = B_c$ for a
qualitative change of the resistivity at {\it zero} temperature. More
precisely, based on the experimental data, it was revealed\cite{Exp2}
that the approximate relation for the critical temperature as a function
of $B$ reads $T_{c}(B) \sim \sqrt{B-B_c}$. This relation implies that at
zero temperature the phase transition happens only when the magnetic field
exceeds the threshold value $B = B_{c}$.

It is remarkable that, as was emphasized in Sec.~\ref{sec:mag-field}, the
existence of such a threshold $B_c$ is a robust consequence of the
mechanism of the magnetic catalysis.  As was pointed out there, the value
$B_c$ is directly related to a nonzero charge density $n$ of carriers,
\be 
|eB_{c}| = \frac{2\pi c n}{N_{f}}, 
\label{threshold}
\ee
and this relation is exact.  For example, by taking $B_{c}= 2.6\times
10^{4}$ G, which was obtained in one of the experiments as an (upper)
estimate of a critical value above which the generation of a gap
presumably occurred,\cite{Exp1} we derive the corresponding charge
density (we use $N_{f}=2$)
\be 
n= \frac{|eB_{c}|}{\pi c } = 1.25 \times
10^{-5} \A^{-2}. 
\label{den-estimate} 
\ee
This should be compared with the charge density (per unit area of a layer)
of carriers in the used sample of graphite. By noting that the area per
carbon atom in a layer is $S=\sqrt{3}a^{2}/4$ where the lattice spacing is
$a\approx2.46\A$,\cite{Wallace} we conclude that the density in
Eq.~(\ref{den-estimate}) corresponds to $n\approx 3.3\times 10^{-5}$ units
of charge per atom. While we do not know the exact density of the sample
used, the given estimate is not unlikely.\cite{Kelly}

\begin{figure}
\epsfxsize=8.0cm
\epsffile[90 -5 580 300]{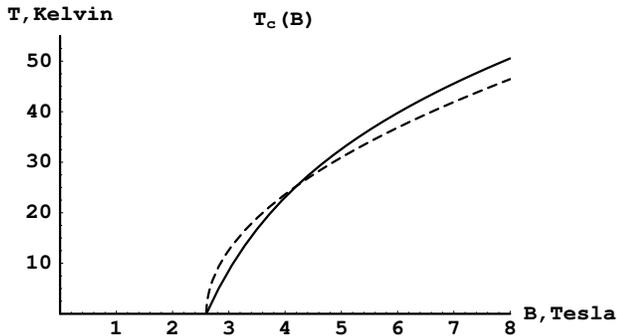}
\caption{The critical temperature as a function of a magnetic field. The
solid line gives the dependence in Eq.~(\ref{T_c}), while the dashed line
corresponds to a dependence with a simple field offset, i.e., $T_{c}\sim
\sqrt{B-B_{c}}$. To plot the figure we used the following parameters:
$N_{f}=2$, $c/v_{F}=375$, $\varepsilon_{0}=2.4$ and $B_{c}=2.6\times
10^{4}$ G.}
\label{Tcrit}
\end{figure}

Notice that the relation $T_{c}(B)  \sim \sqrt{B-B_c} \equiv
\sqrt{1-\nu_{B}} \sqrt{B}$, used in Ref.~\onlinecite{Exp2}, qualitatively
differs from our Eq.~(\ref{T_c}). It is quite remarkable, however, that
the dependence $T_{c}(B)$ in Eq.~(\ref{T_c}) is nearly the same
numerically as the simple square root relation suggested by the
experimental data, see Fig.~\ref{Tcrit}.
Since the relation $|eB_c| = 2\pi c n/N_{f}$ is exact in the dynamics of
the magnetic catalysis, its experimental verification would be a critical
check of the validity of the scenario of the magnetic catalysis in highly
oriented pyrolytic graphite.

Another noticeable experimental observation is that the scale of the
critical temperature is set by the energy distance between the Landau
levels (the Landau scale).\cite{Exp1,SSCom115,Exp2} Therefore, if the
underlying physics of the transition is related to a dynamical generation
of a gap, the typical values of the gap should also be of the same order
as the Landau scale. As we discussed at length in
Sec.~\ref{sec:mag-field}, for the mechanism of the the magnetic catalysis
this fact implies that the pairing dynamics corresponds to the 
strong coupling regime. This in
turn implies that all (or many) Landau levels determine the pairing
dynamics in this case. In connection with that, we would like to point
out that, as the numerical analysis done in Sec.~\ref{condres} shows, the
contribution of higher Landau levels into the conductivity and resistivity
become indeed important for values of the critical temperature $T_c$ of
the order of the Landau scale.

There still remain some unresolved issues in the interpretation of the
experimental data in highly oriented pyrolytic graphite in a magnetic
field.\cite{Exp1,SSCom115,Exp2} The most important of them is the
observation of weak ferromagnetism in that system (for some speculations
concerning its origin see Ref.~\onlinecite {Khvesh}).  We hope to consider
this issue elsewhere.

\section{Conclusion}
\label{conclusion}

In this paper we developed a theory of the magnetic-field-driven
metal-insulator phase transition in planar systems, based on reduced QED.
The general structure of the phase diagram of such systems was established
in two cases, with and without an external magnetic field.  The behavior
of the electric conductivity (resistivity) in these systems was described
in detail. This allowed us to conclude that, in the presence of a magnetic
field, the flavor phase transition in planar systems can indeed manifest
itself as a metal-insulator phase transition in the behavior of the
resistivity $\rho(T,B)$ as a function of the magnetic field and
temperature. It was also shown that there exist clearly distinguishable
signatures of different types of the phase transition. While the
resistivity $\rho(T)$ is a smooth function at the critical point $T = T_c$
in the case of a non-mean-field continuous
phase transition, there are a
discontinuity and a kink in $\rho(T)$ at $T = T_c$ in the cases of the
first order and mean-field phase transitions, respectively.

Based on the experimental data,\cite{Exp1,SSCom115,Exp2} it has been
recently argued that highly oriented pyrolytic graphite shows up a
metal-insulator phase transition, driven by an external magnetic
field.\cite{Khvesh} This might be a nonrelativistic realization of the
phenomenon of the magnetic catalysis originally established in
Refs.~\onlinecite{1,2} in relativistic systems.  In this paper we studied
this possibility rather in detail, elaborating the theory of the magnetic
catalysis in nonrelativistic planar systems and analyzing the temperature
behavior of the resistivity (conductivity) in these systems.  The
conclusion of the present analysis concerning the possibility of this
scenario in highly oriented pyrolytic graphite is quite positive.

One of the central results of this paper is establishing the {\it exact}
relation (\ref{threshold}) for the critical (threshold) value of the
magnetic field at zero temperature in these systems. An experimental
verification of this result would be a crucial test for the present
theory.

Another conclusion of our investigation is that a nonzero magnetic field
alone (even without producing a dynamical gap) can drastically change the
general behavior of the resistivity as a function of temperature. In
particular, in our simplest model with a constant value of the width
parameter, the semiconductor type dependence of the resistivity [i.e.,
$\rho(T)$ decreasing with increasing temperature], seen in the absence of
a magnetic field, can be replaced by a metallic type behavior [i.e.,
$\rho(T)$ increasing with temperature] in the region of not too high
temperatures, when a nonzero field is turned on. In fact, at zero charge
density of carriers, this change of behavior always happens in the range
of temperatures $0<T\lesssim 0.2 v_{F}\sqrt{|eB|/c}$.  This is also seen
at finite but small densities when the diagonal component of conductivity
dominates over the Hall conductivity.

We expect that the results of this paper will be useful for a wide
class of condensed matter planar systems.

\section*{Acknowledgments}
We would like to thank D.~Khveshchenko for numerous valuable discussions
and bringing Ref.~\onlinecite{D-Lee} to our attention. I.A.S. would like
to thank B. Shklovskii for useful discussions. The research of V.P.G. has
been supported in part by the National Science Foundation under Grant No.
PHY-0070986 and by the SCOPES-projects 7~IP~062607 and 7UKPJ062150.00/1 of
Swiss NSF. V.A.M. is grateful for support from the Natural Sciences and
Engineering Research Council of Canada. The work of I.A.S. was supported 
by the U.S. Department of Energy Grant No.~DE-FG02-87ER40328.

\appendix

\section{Symmetry of $(2+1)$-dimensional fermions}
\label{Appfer}

In this Appendix we will consider the symmetry of 4-component fermions on
a plane which carry the flavor index $i=1,2,\dots,N_f$. The three
$4\times4$ $\gamma$-matrices in Eq.~(\ref{L-free}) can be taken to be
\ba
\gamma^0 &=& \left(\begin{array}{cc} \sigma_3 & 0 \\ 
0& -\sigma_3\end{array}\right),\\
\gamma^1 &=& \left(\begin{array}{cc} i\sigma_1&0\\
0&-i\sigma_1\end{array}\right), \\
\gamma^2 &=&\left(\begin{array}{cc}i\sigma_2&0\\
0&-i\sigma_2\end{array}\right).
\label{eq:2}  
\ea
Recall that in $2+1$ dimensions, two sets of matrices
$(\sigma_3,i\sigma_1,i\sigma_2)$ and $(-\sigma_3,-i\sigma_1,-i\sigma_2)$
make inequivalent representations of the Clifford (Dirac) algebra
\begin{equation} \gamma^{\mu} \gamma^{\nu} + \gamma^{\nu} \gamma^{\mu}=
2g^{\mu \nu}, \end{equation} where $\mu, \nu = 0, 1, 2$ and $g^{\mu \nu}=$
diag(1, -1, -1).

There are two matrices,
\begin{equation}
\gamma^3= i\left(\begin{array}{cc} 0& 1 \\ 1& 0\end{array}\right),\quad
\gamma^5= i\left(\begin{array}{cc} 0&1\\-1&0\end{array}\right),
\end{equation}
that anticommute with  $\gamma^0,\gamma^1$ and $\gamma^2$. Therefore for
each four-component spinor, there is a global $U(2)$ symmetry with the
generators
\begin{equation}
I,\quad {1\over i}\gamma^3,\quad \gamma^5,\quad \mbox{and}\quad 
{1\over 2}[\gamma^3,\gamma^5].  
\end{equation}  
Since there are $N_f$ fermion flavors, the full symmetry of the action
(\ref{action}) is 
$U(2N_f)$ with the generators
\begin{equation}
\frac{\lambda^\alpha}{2},\quad 
{\lambda^\alpha\over {2i}}\gamma^3,\quad 
\frac{\lambda^\alpha}{2} \gamma^5,\quad\mbox{and}\quad 
\frac{\lambda^\alpha}{2}{1\over2}[\gamma^3,\gamma^5],
\end{equation}  
where $\lambda^\alpha/2$, with $\alpha=0,1,\dots, N_f^2-1$, are
$N^{2}_{f}$ generators of $U(N_{f})$. 

Adding a mass (gap) term $\Delta_{0}\bar\psi\psi$ into the action
(\ref{action}) would reduce the $U(2N_{f})$ symmetry down to the
$U(N_f)\times U(N_f)$ with the generators
\be
\frac{\lambda^\alpha}{2},\quad
\frac{\lambda^\alpha}{2}{1\over2}[\gamma^3,\gamma^5],
\ee  
with $\alpha=0,1,\dots, N_f^2-1$. This implies that the dynamical 
generation of the fermion gap leads to the spontaneous breakdown of 
the $U(2N_f)$ down to the $U(N_f)\times U(N_f)$.

\section{Derivation of polarization function and gap equation}
\label{AppA}

In this Appendix, we give the details of the calculations of the time
component of the gauge field polarization function, as well as the
derivation of the gap equation at finite chemical potential and finite
temperature. We will consider only the case of zero magnetic field.  The
polarization function and the gap equation in $(2+1)$-dimensional QED with
an external magnetic field were given in Ref.~\onlinecite{Shpagin} where
the method of Ref.~\onlinecite{2} was used.

\subsection{Polarization function}

The general expression of the time component of the vacuum 
polarization function is given by\cite{footnote1}
\ba
\Pi(\Omega_{m}, \vec{p}) &=& \frac{2\pi}{\varepsilon_{0}}
e^2TN_f \sum_{n=-\infty}^{+\infty} \int
\frac{d^2k}{(2\pi)^2} \nonumber \\
&\times& \mbox{tr}
\left[\gamma_0 S(\Omega_{m}+\omega_n, \vec{p}+\vec{k})
\gamma_0 S(\omega_n, \vec{k})\right],
\label{Pi-gen}
\ea
where $S(\omega_n, \vec{k})$ is the fermionic quasiparticle propagator 
whose explicit form reads  
\be
S(\omega_n, \vec{k})= \frac{i}{(i\omega_n - \mu)\gamma_0 
+ (\vec{k}\cdot\vec{\gamma}) + \Delta_{T}(\mu)}.
\ee
In Eq.~(\ref{Pi-gen}), the Matsubara frequencies are denoted by $\omega_n
\equiv (2n+1)\pi T $ and $\Omega_m \equiv 2 m\pi T$. Also notice that the
expression on the right hand side is multiplied by an additional factor
$2\pi/\varepsilon_{0}$, in accordance with our definition of the polarization
function. After taking the trace over the Dirac indices and using the
Feynman parametrization, we obtain
\ba
&&\Pi(0, \vec{p}) = \frac{8\pi}{\varepsilon_{0}}
e^2TN_f \sum_{n=-\infty}^{+\infty} \int_0^1 dx \int
\frac{d^2k}{(2\pi)^2}\nonumber \\
&\times& 
\Bigg[ \frac{1}{(\omega_n + i\mu)^2 + v_F^2p^2x(1-x) 
+ v_F^2k^2 + \Delta_{T}^2(\mu)}  
\nonumber \\ 
&-&\frac{2[v_F^2k^2 + \Delta_{T}^2(\mu)]}{[(\omega_n + i\mu)^2 
+ v_F^2p^2x(1-x) + v_F^2k^2 + \Delta_{T}^2(\mu)]^2} \Bigg].
\ea
By calculating the sum over $n$, we get
\begin{eqnarray}
&&\Pi(0, \vec{p}) = \frac{e^2N_f}{2\varepsilon_{0}} 
\int_0^1 dx \int_0^{\infty} \frac{dk^2}{Y^2}
\Bigg[ \frac{v_F^2p^2x(1-x)}{Y} \nonumber \\
&\times&\tanh\frac{Y+\mu}{2T}
+\frac{v_F^2k^2 + \Delta_{T}^2(\mu)}{2T\cosh^2\frac{Y+\mu}{2T}} 
+ (\mu \to -\mu) \Bigg],
\end{eqnarray}
where $Y=\sqrt{v_F^2k^2 + v_F^2p^2x(1-x) + \Delta_{T}^2(\mu)}$. 
By changing the integration variable, $k^2 \to Y$, and integrating
by parts, we finally arrive at the following convenient representation:
\ba
\Pi(0, \vec{p}) &=& \frac{2Te^2N_f}{\varepsilon_{0} v_F^2} \int_0^1 dx 
\Bigg[ \ln \left( 2 \cosh\frac{R_x + \mu}{2T}\right) \nonumber \\ 
&-&\frac{\Delta_{T}^2(\mu)}
{2TR_x}\tanh\frac{R_x+\mu}{2T}+ (\mu \to -\mu)\Bigg],
\ea
where $R_x=\sqrt{v_F^2p^2x(1-x)+\Delta_{T}^2(\mu)}$.

\subsection{Gap equation}
\label{gapB}

The general Schwinger-Dyson (gap) equation for the quasiparticle 
propagator reads
\ba
S^{-1}(\omega_{m},\vec{p}) &=& S_{0}^{-1}(\omega_{m},\vec{p}) 
-T\sum_{n=-\infty}^{\infty} \int \frac{d^2k}{(2\pi)^2}
\nonumber \\
&\times &
\gamma^{0} S(\omega_{n},\vec{k}) \gamma^{0} U(\vec{p}-\vec{k}).
\ea
By neglecting the wave function renormalization,\cite{footnote}
we derive the following gap equation:
\ba
\Delta(p) &=& \frac{e^2T}{2\pi \varepsilon_{0}} 
\sum_{n=-\infty}^{+\infty} \int
\frac{\Delta(k) d^2k}{(\omega_n + i\mu)^2 + v_F^2k^2 +
\Delta_{T}^2(\mu)} \nonumber \\
&\times & \frac{1}{|\vec{p}-\vec{k}| + \Pi(0,\vec{p}-\vec{k})},
\label{B7}
\ea
where $\Delta_{T}(\mu)\equiv \Delta(p)|_{p=0}$.
Here the interaction is taken in the so-called instantaneous exchange
approximation. This means that the retardation effects of the gauge 
field are neglected which is justified in a nonrelativistic model.

By neglecting the dependence of the gap on the Matsubara frequency, 
we could perform the sum over $n$ explicitly. Then, the result reads
\ba
\Delta(p) &=& \frac{\pi e^2}{\varepsilon_{0}} \int
\frac{d^2k}{(2\pi)^2} \frac{\Delta(k)}{E_{k}} 
\frac{\sinh\frac{E_{k}}{T}}{\cosh\frac{E_{k}}{T} + \cosh\frac{\mu}{T}}
\nonumber \\
&\times&
\frac{1}{|\vec{p}-\vec{k}| + \Pi(0,\vec{p}-\vec{k})},
\label{gap-A7}
\ea
where $E_{k}=\sqrt{v_F^2k^2 + \Delta_{T}^2(\mu)}$. By using the standard 
approximation for the kernel of the integral equation,
$f(|\vec{p}-\vec{k}|) \to f(p)\theta(p-k) + f(k)\theta(k-p)$, 
we obtain the following gap equation:
\ba
\Delta(p) &=& \frac{e^2}{2\varepsilon_{0}v_F} 
\int_{\epsilon}^{\Lambda} dk \Delta(k)
\frac{\sinh\frac{v_Fk}{T}}{\cosh\frac{v_Fk}{T} + \cosh\frac{\mu}{T}}
\nonumber \\
&\times&
\left[\frac{\theta(p-k)}{p +\Pi(0,\vec{p})} 
+\frac{\theta(k-p)}{k + \Pi(0,\vec{k})}\right],
\label{gap-eq-A9}
\ea
where the infrared cutoff $\epsilon$ is given by a larger value of
$\Delta_{T}(\mu)/v_F$ or $\sqrt{\mu^2 - \Delta_{T}^{2}(\mu)}/v_F$, and
where we also utilized the bifurcation method in which a nonlinear gap
equation is replaced by a linear approximation (compare with the
discussion in subsection \ref{dyngap}). This is achieved by substituting
the trivial value of the gap in $E_{k}$ and introducing an infrared cutoff
in the integral on the right hand side of Eq.~(\ref{gap-A7}).

\section{Derivation of effective potential at $\mu\neq 0$}
\label{AppB}

In this appendix, we will construct the effective potential of the
composite field $\sigma=-\langle \bar{\psi} \psi \rangle$ by using the
method of Ref.~\onlinecite{Mir}. For the purposes of this paper, it is
sufficient to consider only the case of a nonzero
chemical potential. The generalization to some other cases (for example,
with an external magnetic field) is also possible, see for example
Ref.~\onlinecite{MLNS}. 

In order to derive the effective potential as a function of the composite
field $\sigma$, one should introduce a term with a constant external
source $J$ coupled to the corresponding composite operator in the action,
and construct the generating functional $W(J)$. The effective potential,
then, is defined through the Legendre transform as follows:\cite{Mir}
\be
V(\sigma) = -w(J)+ J \sigma = \int^{\sigma} d\sigma J(\sigma), 
\label{eff-pot-def}
\ee
where $\sigma=\partial w(J)/\partial J$, $w(J)\equiv W(J)/V_{2+1}$, 
and $V_{2+1}$ is the space-time volume. In the last expression, 
the source $J$ should be regarded as a function of the field $\sigma$.

The effect of the external source $J$ could be easily taken into account
in the gap equation (\ref{gap-eq-n}): one should simply replace
$\Delta_{p} \to \Delta_{p} -J$ on the left hand side of the equation.
Then, the solution to the equation, satisfying the infrared boundary
condition, takes the following form:
\be
\Delta_{p} =\frac{\Delta}{\sin\delta}\sqrt{\frac{\epsilon}{p}}
\sin\left[\frac{\nu}{2}\ln\frac{p}{\epsilon}+\delta\right],
\ee
where $\nu = \sqrt{4\lambda-1}$, $\epsilon = \mbox{max}
\{\Delta/v_{F}, \sqrt{\mu^2-\Delta^2}/v_{F}\}$, and $\delta = \arctan\nu$.
The overall normalization of the above solution is fixed by choosing
$\Delta_{p=\epsilon} = \Delta$.  The ultraviolet boundary condition,
\be
J=\left.(\Delta_{p}+ p\Delta_{p}^{\prime})\right|_{p=\Lambda},
\ee
on the other hand, produces the relation:
\be
J=\frac{\Delta}{\sin(2\delta)}\sqrt{\frac{\epsilon}{\Lambda}}
\sin\left[\frac{\nu}{2}\ln\frac{\Lambda}{\epsilon}+2\delta\right].
\ee
As it should be, the equation for the dynamical gap $\Delta_{0}$ is
obtained in the limit of vanishing source $J=0$. At zero chemical
potential, in particular, the equation for the gap takes the form:
\be
\frac{\nu}{2}\ln\frac{\Lambda v_{F}}{\Delta_0}=\pi-2\delta.
\label{gap-0}
\ee
For the derivation of the effective potential, we also need to know the
expression for the field $\sigma$. By definition, it is equal to the trace
of the fermion propagator. Thus, we get
\ba
\sigma &=& -\langle\bar\psi\psi\rangle=-\left.
\frac{N_f}{\pi\lambda v_{F}}
p^2\Delta^\prime(p)\right|_{p=\Lambda} \nonumber \\
&=&\frac{N_f\Delta\sqrt{\epsilon\Lambda}}
{\pi\lambda v_{F}\sin(2\delta)}
\sin\left[\frac{\nu}{2}\ln\frac{\Lambda}{\epsilon}\right].
\ea 
Now, by making use of Eq.~(\ref{gap-0}), we trade the cutoff parameter
$\Lambda$ for $\Delta_{0}$. After that, we derive the following
approximate relations for the case of small $\nu$ we are interested in:
\ba
J(\Delta) &\simeq & -\frac{\Delta}{4}
\sqrt{\frac{\epsilon}{\Lambda}}
\ln\frac{\Delta_0}{\epsilon v_{F}}, \\
\sigma(\Delta)&\simeq &\frac{N_f\Delta\sqrt{\epsilon\Lambda}}{\pi v_{F}}
\left(4-\ln\frac{\Delta_0}{\epsilon v_{F}}\right).
\label{sigma}
\ea 
As will become clear in a moment, these two expressions contain all the
information needed for reconstructing the potential.  Indeed, the
definition of the effective potential in Eq.~(\ref{eff-pot-def}) can be
rewritten as follows:
\be
V(\sigma)=\int\limits^\Delta
d\Delta\frac{d\sigma(\Delta)}{d\Delta}J(\Delta)+f(\mu),
\ee
where the most general integration constant $f(\mu)$ was 
added on the right hand side. This new representation leads to the 
final result, 
\ba
V(\Delta)&=&\frac{N_f \Delta^2 \sqrt{\mu^2-\Delta^2}}
{2\pi v_{F}^{2}}\Bigg[
\frac{1}{4}\ln^2\frac{\sqrt{\mu^2-\Delta^2}}{\Delta_0} \nonumber \\
&+&\ln\frac{\sqrt{\mu^2-\Delta^2}}{\Delta_0}
-\frac{2\mu^2+\Delta^2}{3\Delta^2}\Bigg]
+f_{1}(\mu),
\label{second_branch}
\ea
for $\Delta\leq |\mu|/\sqrt{2}$, and
\ba
V(\Delta) &=& \frac{N_f\Delta^3}{2\pi v_{F}^{2}}\left[
\frac{1}{4}\ln^2\frac{\Delta}{\Delta_0}
+\ln\frac{\Delta}{\Delta_0}-\frac{1}{3}\right]
+f_{2}(\mu),
\label{first_branch}
\ea
for $\Delta\geq |\mu|/\sqrt{2}$. In these equations we used the freedom of
choosing the integration constants in expressions (\ref{second_branch})
and (\ref{first_branch}) as follows:
\ba
f_{1}(\mu) &=& \frac{\sqrt{2} N_f |\mu|^3}{6\pi v_{F}^{2}}
+f_{2}(\mu) ,\\
f_{2}(\mu) &=&-\frac{N_f}{6\pi v_{F}^{2}}
\left(|\mu|-\Delta_{0}\right)^{2} \left(|\mu|+2\Delta_{0}\right)
\nonumber\\
&&\times \theta(|\mu|-\Delta_{0})\theta(\mu_{c}-|\mu|)
\nonumber\\
&-&\frac{N_f}{6\pi v_{F}^{2}}
\left[(\sqrt{2}-1)(|\mu|^{3}-\mu_{c}^{3}) \right. \nonumber \\
&&\left.
+\left(\mu_{c}-\Delta_{0}\right)^{2} \left(\mu_{c}+2\Delta_{0}\right)
\right]\theta(|\mu|-\mu_{c}).
\ea
This choice insures that the potential is continuous at the matching 
point $\Delta = |\mu|/\sqrt{2}$, and that it is normalized so that its
partial derivative with respect to the chemical potential at the global 
minimum is equal (up to a sign) to the charge density:
\be
\frac{\partial V(\Delta_{0},\mu)}{\partial \mu}
\equiv \frac{N_{f}(\Delta_{0}^{2}-\mu^{2})}{2\pi v_{F}^{2}}
\mbox{sgn}(\mu)\theta(|\mu|-\Delta_{0}),
\label{dv/dmu}
\ee
for $|\mu| < \mu_{c}$, and 
\be
\frac{\partial V(0,\mu)}{\partial \mu}
\equiv -\frac{N_{f}\mu^{2}}{2\pi v_{F}^{2}}\mbox{sgn}(\mu),
\ee
for $|\mu| >\mu_{c}$. Here we used the expression for the charge 
density in Eq.~(\ref{density-B0-T0}).

Now, the effective potential as a function of the compo\-site field
$\sigma$ is defined parametrically through Eqs.~(\ref{second_branch}),
(\ref{first_branch}) and (\ref{sigma}). This dependence is graphically
shown in Fig.~\ref{fig:eff-pot} for a few different values of the chemical
potential. As is clear from the figure, the presence of a nonzero chemical
potential considerably changes the behavior of the effective potential. In
particular, a new local minimum develops at the origin and its depth
gradually increases with $\mu$. The competition of the two minima, located
at $\sigma=0$ and $\sigma_{0}\equiv \sigma(\Delta_0)$, results in a first
order phase transition. Such a transition happens when the depths of
effective potential at its two minima become equal. By making use of this
criterion, we derive the analytical expression for the critical value of
the chemical potential,
\be 
\mu_{c}=\frac{\Delta_0}{(2-\sqrt{2})^{1/3}}\simeq 1.195\Delta_0.
\ee


\begin{references}
\bibitem{footnoteN} The symmetry $U(2N_f)$ can be also explicitly
broken by additional local interactions (e.g., four-fermion ones) 
possessing a lower symmetry.
\bibitem{footnote} The wave function renormalization is a next
to the leading order effect in the $1/N_{f}$ expansion. The experience
of working with the gap equation in QED$_{2+1}$ suggests that neglecting
it does not affect results too much even for values of $N_{f}$ of order
one. For example, in the Landau gauge, the critical value for the number
of fermion flavors is $N_{f}^{cr} \simeq 3.24$ and $N_{f}^{cr} \simeq
4.32$ in the cases of neglecting the wave function renormalization and
taking it into account, respectively.\cite{QED_{2+1}}
\bibitem{footnote1} We use the symbols $G$ and $S$ for the
fermion propagator in the cases with and without magnetic
field, respectively.
\end{references}
\end{document}